\documentclass{sig-alternate}
\usepackage[lined,boxed,commentsnumbered]{algorithm2e}
\usepackage{array}
\usepackage{amssymb}
\usepackage{url}
\usepackage{cite}
\usepackage{rotating}

\newcommand{\gridsize}{0.23}
\newtheorem{teorema}{\textsc{Theorem}}

\newtheorem{proposition}{\textsc{Proposition}}
\newtheorem{corolari}{\textsc{Corollary}}

\newcommand{\smallSpace}{\vspace*{-0.1cm}}

\newcommand{\proposicioConjunts}
{
	Let $G=(V,E)$ be a graph and $\emptyset\ne S\subseteq V$. Then,
	\begin{enumerate}
	\item[\emph{(i)}] $0 \leq WCC(S) \leq 1$.
	\item[\emph{(ii)}] {\it $WCC(S)=0$ if and only if $S$ has no triangles.}
	\item[\emph{(iii)}]{\it $WCC(S)=1$ if and only if $S$ is a clique
		with  $vt(x,V\setminus S) = 0$ for all $x \in S$.}
	\end{enumerate}
}

\newcommand{\proposicionsNode}
{
	Let $G=(V,E)$ be a graph and $\emptyset\ne S\subseteq V$. Then, 
		\begin{enumerate}
	\item[\emph{(i)}] $0\le WCC(x,S)\le 1$ for all $x\in V$.
	\item[\emph{(ii)}] $WCC(x,S)=0$ if and only if $t(x,S)=0$.
	\item[\emph{(iii)}] $WCC(x,S)=1$ if and only if   
		$vt(x,S)= |S\setminus\{x\}|\ge 2$, and $vt(x,V\setminus S)=0$.
\end{enumerate}

}

\newcommand{\teoremaBridgeEdge}
{
	Let $S_{1}$ and $S_{2}$ be two communities in a partition of 
	graph $G=(V,E)$ such that:
	\begin{enumerate}
		\item[\emph{(i)}] $S_{1}$ and $S_{2}$ are the set of vertices of two different connected components.
		\item[\emph{(ii)}] $WCC(S_{1})>0$.
	\end{enumerate}
	Then, the following inequality holds:$$WCC(\{S_{1},S_{2}\})>WCC(\{S_{1}\cup S_{2}\}).$$
}

\newcommand{\teoremaCliqueNode}
{
Let $G=(V,E)$ be a random graph of order $r$
in which each edge occurs independently with probability $p$. Let $v\not\in V$ be a vertex 
adjacent to $d\ge 2$ vertices of  $V$. Consider
  the two partitions $\mathcal{P}_1=\{V \cup \{v\}\}$ and
  $\mathcal{P}_2=\{V,\{v\}\}$. Then,
 \begin{enumerate}
 \item[\emph{(i)}] $\displaystyle(r+1)WCC(\mathcal{P}_1)=(r-1)p+2\cdot d \cdot r^{-1}$.
 \item[\emph{(ii)}]
$\displaystyle (r+1)WCC(\mathcal{P}_2)
       =(r-d)p$\newline
        $\displaystyle\phantom{(r+1)WCC(\mathcal{P}_2=}+\frac{d}{r}
        \cdot\frac{((r-1)p+1)(r-1)(r-2)p^2}{(r-1)(r-2)p^2+2(d-1)}$.
\item[\emph{(iii)}] For large enough $r$, $WCC(\mathcal{P}_1)>WCC(\mathcal{P}_2)$ if
and only if $d>rp\left(\sqrt{p^2+2p+9}-(1+p)\right)/4$. 
\end{enumerate}
}

\newcommand{\corolariClique}
{
Let $S$ be a clique of order $r$. Given a vertex $v$, there must exist at least $0.37\cdot r$ edges between $v$ and $S$
to hold $WCC(S \cup \{v\}) > WCC(S,\{v\})$.
}

\newcommand{\teoremaCliqueClique}
{
Let $G=(V,E)$ be a graph of order $n$ which consists
  of two cliques $K_r$ and $K_s$ of orders $r$ and $s$, respectively,
  that intersect in a vertex t. Assume $r\ge s\ge
  4$.
  {
\begin{enumerate}
\item[\emph{(i)}] If $\mathcal{P}_1=\{K_r\cup K_s\}$, then
\begin{equation}
\label{f1}
n\cdot WCC(\mathcal{P}_1)=\frac{(r-1)(r-1)}{r+s-2}+\frac{1}{r+s-2}+\frac{(s-1)(s-1)}{r+s-2};
\end{equation}
\item[\emph{(ii)}] if $\mathcal{P}_2=\{K_r, \ K_s \setminus \{t\}\}$, then
\begin{align}
n\cdot WCC(\mathcal{P}_2)=(r-1)
&+\frac{(r-1)(r-2)}{(r-1)(r-2)+(s-1)(s-2)} \nonumber \\
&+\frac{(s-1)(s-2)(s-3)}{(s-1)(s-2)}; \label{f2} 
\end{align}
\item[\emph{(iii)}] if 
$\mathcal{P}_3=\{K_r\setminus \{t\}, \{t\},\  K_s\setminus \{t\}$, then 
\begin{align}
n\cdot WCC(\mathcal{P}_3)=
&\phantom{+}\frac{(r-1)(r-2)(r-3)}{(r-1)(r-2)}\\
&+\frac{(s-1)(s-2)(s-3)}{(s-1)(s-2)};\label{f3}
\end{align}
\item[\emph{(iv)}] 
$WCC(\mathcal{P}_3)\}\le WCC(\mathcal{P}_2)$.
\item[\emph{(v)}] 
$\max\{WCC(\mathcal{P}_1),WCC(\mathcal{P}_2),WCC(\mathcal{P}_3)\}=
WCC(\mathcal{P}_2)$. 
\end{enumerate}
  }

}

\begin{document}

\conferenceinfo{CIKM'12,} {October 29--November 2, 2012, Maui, HI, USA.}
\CopyrightYear{2012}
\crdata{978-1-4503-1156-4/12/10}
\clubpenalty=10000
\widowpenalty = 10000

\title{Shaping Communities out of Triangles}

\numberofauthors{3}
\author{
\alignauthor
Arnau Prat-P\'erez\\
      \affaddr{DAMA-UPC}\\
      \affaddr{Universitat Polit\`ecnica de Catalunya}\\
      \email{aprat@ac.upc.edu}
\alignauthor
David Dominguez-Sal\\
      \affaddr{DAMA-UPC}\\
      \affaddr{Universitat Polit\`ecnica de Catalunya}\\
      \email{ddomings@ac.upc.edu}
\alignauthor
Josep M.Brunat\\
      \affaddr{Departament Matem\`atica Aplicada II}\\
      \affaddr{Universitat Polit\`ecnica de Catalunya}\\
      \email{josep.m.brunat@upc.edu}
\and
\alignauthor
Josep-LLuis Larriba-Pey\\
      \affaddr{DAMA-UPC}\\
      \affaddr{Universitat Polit\`ecnica de Catalunya}\\
      \email{larri@ac.upc.edu}
}


\maketitle

\begin{abstract}

Community detection has arisen as one of the most relevant topics in the field
of graph data mining due to its importance in many fields such as biology,
social networks or network traffic analysis.
The metrics proposed to shape communities are
generic and follow two approaches: maximizing the internal 
density of such communities or
reducing the connectivity of the internal vertices 
with those outside the community.
However, these metrics take the edges as a set and 
do not consider the internal layout of the edges in the community.
We define a set of properties oriented to social networks that ensure that
communities are cohesive, structured and well defined. Then,
we propose the Weighted Community Clustering ($WCC$), which is a community
metric based on triangles. We 
proof that analyzing communities by triangles gives
communities that fulfill the listed set of properties, in contrast to
previous metrics.
Finally, we experimentally show that WCC correctly captures
the concept of community in social networks using real and syntethic datasets, 
and compare statistically some of the most relevant community detection 
algorithms in the state of the art.
\end{abstract}

\section{Introduction}

Although graphs are a very intuitive representation of many datasets, retrieving
information from them is far from easy. The increasingly growing
datasets during the last years have made it very difficult to intuitively
extract and analyze the information of the graphs generated from 
those data sources. Large graphs have often many
relationships that make their visual analysis impossible and make the 
understanding of the structural components of the graph difficult.

Communities are informally defined as sets of vertices which are densely 
connected but scarcely connected to the rest of the graph.
The retrieval
of vertex communities (or clusters) provides information
about the sets of vertices that respond to a similar
concept~\cite{girvan2002community}. In social networks, communities identify
groups of users with similar interests, locations, friends or occupations. 
This information is useful to craft new
ways to represent data in visual analysis applications~\cite{di2007graph}
or to reduce the access times to this data thanks to a more coalesced
data placement~\cite{prat2011social}.

Several metrics have been proposed as indicators of the quality of a community
~\cite{newman2004finding,kannan2004clusterings,leskovec2010empirical}.
Among them, modularity and conductance are those which have 
become more popular~\cite{fortunato2010community}
and precise~\cite{leskovec2010empirical}, respectively.
Modularity compares the internal edge density of the community with the
average edge density of the graph. On the other hand, conductance computes
the ratio between edges inside the community and the edges in the 
frontier (i.e.~the cut) of the community. Both metrics take the edges as a set
of objects without paying attention to the internal structure of the community. 
One of the consequences is that the optimization of such metrics 
generates communities without noticeable structure that empirically
cannot sometimes be considered communities~\cite{leskovec2010empirical}.

We find that the informal definition of community stated previously is too lax
for social networks, because it does not consider the internal
edge layout of the community. 
As a first contribution, we introduce a set of basic structural properties
that a good community metric for social networks must fulfill.
These properties ensure that communities are cohesive, structured and well
defined. 
An example of these properties is that communities in social networks
must be dense in terms of triangles. 
A triangle is a 
transitive relation between three
vertices. For example, a triangle appears when $A$ is a friend of $B$, $B$ is
a friend of $C$, and $C$ is a friend of $A$. 
Social networks are known to contain more triangles 
than expected by chance (Erd\"os-R\'enyi graph), 
which gives a community structure 
to the graph~\cite{newman2003social,newman2001structure,
shi2007networks,satuluri2011local}. The triangle is a simple structure that
depicts a strong relation among three vertices. Furthermore, complex dense 
structures such as a clique contain a large number of triangles.
Another example is the absence of bridges
in a community. A bridge is an edge 
that connects two connected components, and hence, having 
the two connected components as two different communities 
is more natural and intuitive~\cite{fortunato2007resolution}.
Surprisingly, such requirements are not met by state of the 
art metrics, so they do not provide satisfactory communities.

As a second contribution, we design a community detection metric
called \textit{Weighted Community Clustering} ($WCC$). $WCC$ is based 
on the notion that triangles are a good indicator of community structure. 
$WCC$ takes into account the density and the layout of the triangles 
to rate the quality of a community. We also prove that
our triangle based approach fulfills the introduced properties.

Finally, we show experimentaly that there is a correlation between communities
with good $WCC$ value and desirable statistical values. 
We show that while communities with
large $WCC$ are cohesive and dense, others with good modularity and conductance
values are not. We also compare the most used algorithms in the state of
the art using $WCC$.


The paper is structured as follows: in Section~\ref{section:related-work}, we
review the state of the art.
In Section~\ref{section:wcc}, we introduce the problem of community detection,
propose the new metric ($WCC$) and introduce the properties. In
Section~\ref{section:comparison}, we show that the current metrics in the state
of the art do not fulfill the properties proposed. Finally, in
Section~\ref{section:experiments}, we compare several community detection
methods using the $WCC$ and in Section~\ref{section:conclusions} we give
guidelines for future work and conclusions.

\section{Related Work}\label{section:related-work}

There are basically two types of metrics to evaluate the
quality of a community. First, those that focus on the
internal density of the community. The most widely used metric that falls into
this
category is the {\it modularity}, which was proposed 
by Newman et al.\cite{newman2004finding}.
Modularity measures the
internal connectivity of the community (omitting the external connectivity) 
compared to an Erd\"os-R\'enyi graph model. It has become very popular in the
literature, and
a lot of algorithms are based on maximizing it. Algorithms apply several 
optimization procedures: agglomerative greedy~\cite{clauset2004finding},
simulated annealing strategy~\cite{medus2005detection}
or multistep approaches~\cite{blondel2008fast}.

However, it has been reported that modularity has resolution
limits~\cite{fortunato2007resolution,good2010performance}.
Communities detected by modularity depend on the total graph size, and thus
for large graphs, small well defined communities are never found.
This means that maximizing the modularity
leads to partitions where communities are far from intuitive. This is illustrated
in Figure~\ref{fig:modularity_resolution} by an example.

The second type of metrics consists of those that focus on reducing the number of 
edges connecting communities. In~\cite{kannan2004clusterings}, 
Kannan et al. introduce the conductance. 
Conductance, is the ratio between the edges going outside the community 
and the total number of edges between members of the 
community. However, conductance suffers
from the fact that for any graph, the partition 
with a unique community containing
all the vertices of the graph obtains the best conductance, 
making its direct optimization not viable. 
A recent survey~\cite{leskovec2010empirical} of community metrics
discusses the performance of many metrics on real networks: the cut
ratio~\cite{fortunato2010community}, 
the normalized cut~\cite{shi2000normalized}, the Maximum-ODF (Out Degree
Fraction), the Average-ODF and Flake-ODF~\cite{flake2000efficient}. 
In this survey, Leskovec et al. showed that, among all 
these metrics, conductance is the metric that best captures 
the concept of community. Furthermore,
their results reveal that the quality of communities decreases significantly
for those of size greater than around 100 elements.

\section{Weighted Community Clustering}~\label{section:wcc}
\subsection{Problem Formalization}\label{section:formalization}
Given a graph $G=(V,E)$, the problem is to classify the vertices of
the graph into disjoint cohesive sets. The criterion to measure
the cohesion of the sets is formally obtained by defining a metric, that is, a
function $f$ that assigns to each subset $S$ of $V$ a real number
such that $0\le f(S)\le 1$. A \emph{community} is a set of vertices $S$, 
on which we compute a degree of cohesion $f(S)$.
Good communities have a large $f(S)$ and bad communities have a small $f(S)$.
The adequate metric $f$ for a given context 
(social networks, biology, etc.) captures the features of
the communities in that context.

A \emph{partition} of $V$
is a set $\mathcal{P}=\{C_1,\ldots,C_n\}$ of
non-empty pairwise disjoint subsets of $V$ such that
$C_1\cup\cdots\cup C_n=V$. A metric $f$ in $G$ defines in a natural way
a value $f(\mathcal{P})$ in each partition $\mathcal{P}$ of $V$ by taking the 
weighted average of the value of the function on the sets of the partition:
\begin{equation}
f(\mathcal{P})=\frac{1}{|V|}\sum_{i=1}^n \left( |C_i|\cdot f(C_i) \right) .
\label{equation:f_partition}
\end{equation}
For a given graph and a given metric $f$ in $G$, the goal is to obtain
an \emph{optimal partition}, that is, a partition $\mathcal{P}$ such
that $f(\mathcal{P})$ is maximum. We call 
the communities in an optimal partition 
the \emph{optimal communities} of the graph.

\subsection{Metric Definition}\label{section:metric}

A natural way to define the cohesion of a community is 
to define first the degree of cohesion of a vertex $x$ with respect to a set $S$.
That is, a function $f$ that assigns to the pair $(x,S)$ 
a real number $f(x,S)$ in the range $0\le f(x,S)\le 1$. 
Then the metric on $S$ is defined by taking the average of $f(x,S)$ with
$x\in S$, that is,
\begin{equation}
f(S)=\frac{1}{|S|}\sum_{x\in S}f(x,S).
\label{equation:f}
\end{equation}
In this paper, we propose a definition of metric $f(S)$ 
that we call \emph{Weighted Community Clustering} ($WCC$), 
which computes the level of cohesion of a set of vertices $S$.
In the rest of the paper, we refer to our proposal for $f(S)$ as $WCC(S)$.

In order to define $WCC(S)$ we start defining $WCC(x,S)$. 
With that objective, we denote by $t(x,S)$ the number of triangles that 
vertex $x$ closes with vertices in $S$ and by $vt(x,S)$ 
the number of vertices of $S$ that form at least one triangle with $x$. 
$WCC(x,S)$ is calculated as follows:
\begin{equation}
WCC(x,S) = \left\{\begin{array}{ll}
\frac{t(x,S)}{t(x,V)} \cdot \frac{vt(x,V)}{|S \setminus \{x\}|+vt(x,V \setminus S)} & \text{if $t(x,V) \neq 0$};\\
0 & \text{if $t(x,V)=0$}.
\end{array}\right.
\label{eq:ccc_normalizat2}
\end{equation}
Note that $|S\setminus x|+vt(x,V\setminus S) = 0$ implies that $S=\{x\}$ and $vt(x,V)=0$. Then the condition 
$|S\setminus x|+vt(x,V\setminus S)=0$ is included in the condition $t(x,V)=0$.


The left fraction of $WCC(x,S)$ is the ratio of triangles that vertex 
$x$ closes with set $S$, as opposed
to the number of triangles that $x$ closes with the whole graph. 
On the other hand, the right fraction is
the number of vertices that close at least one triangle with $x$,
with respect to the union of such set and $S$.

The cohesion of a partition is computed 
as stated in Equation~(\ref{equation:f_partition}),
by using $WCC(S)$. Therefore, an optimal partition is such that,
for all vertices of the graph, the two 
factors of Equation~(\ref{eq:ccc_normalizat2}) are optimized.
The left fraction is maximized for a vertex $x$ when set $S$ includes \emph{all}
the vertices that form triangles with $x$. 
Note that since a pair of vertices can build many triangles, the left term
rewards including the vertices that build more triangles with $x$. 
The right fraction is 
maximized for $x$ when set $S$ contains \emph{no} vertices such that $x$ 
does not form triangles. The maximization process is a 
compromise between both terms:
the left term is optimized by including additional vertices in the set, 
but the second is optimized by removing vertices from the set. This behavior implies that
good communities are those with a significant number of triangles well distributed among all the vertices.

Proposition~\ref{proposition:proposicionsNode} introduces a set of natural properties of $WCC(x,S)$ (proofs
are available in Appendix~\ref{appendix:formal_proof_init}).
\begin{proposition}
\proposicionsNode
\label{proposition:proposicionsNode}
\end{proposition}
The value of $WCC(x,S)$  indicates the fitness 
of vertex $x$ to become part of the set of vertices $S$. This 
value is a real number between 0 and 1 
(Proposition~\ref{proposition:proposicionsNode}~(i)). These two extremes
are only seen in particular 
situations (Proposition~\ref{proposition:proposicionsNode}~(ii-iii)).
For a given vertex $x$, in order to have
some degree of cohesion with a $S$, the vertex
must at least form one triangle with two other vertices in set $S$.
If a vertex builds no triangle with the vertices in $S$, 
then the cohesion of the vertex with respect to the set is zero.
On the other hand, the value one is reached 
if and only if all the vertices of $S$ form at least one triangle with $x$. 
This property reflects the fact that
the cohesion of a vertex $x$ with respect to a set $S$, is maximized when $S$ includes exactly all the vertices
that close triangles with $x$. 
Furthermore, from the
point of view of the $WCC$, only those edges in $E$ closing 
at least one triangle are relevant and influence the cohesion of a vertex.

$WCC(S)$ indicates the quality of a community. We infer several properties on $WCC(S)$ from Proposition~\ref{proposition:proposicionsNode}  
(see proofs in Appendix~\ref{appendix:formal_proof_B}).

\begin{proposition}
\proposicioConjunts
\end{proposition}

\begin{figure}[t!]

\begin{minipage}{1\linewidth}
\textbf{\small Example: }
\end{minipage}	
\begin{minipage}{0.22\linewidth}
\centering
\includegraphics[width=0.6\linewidth]{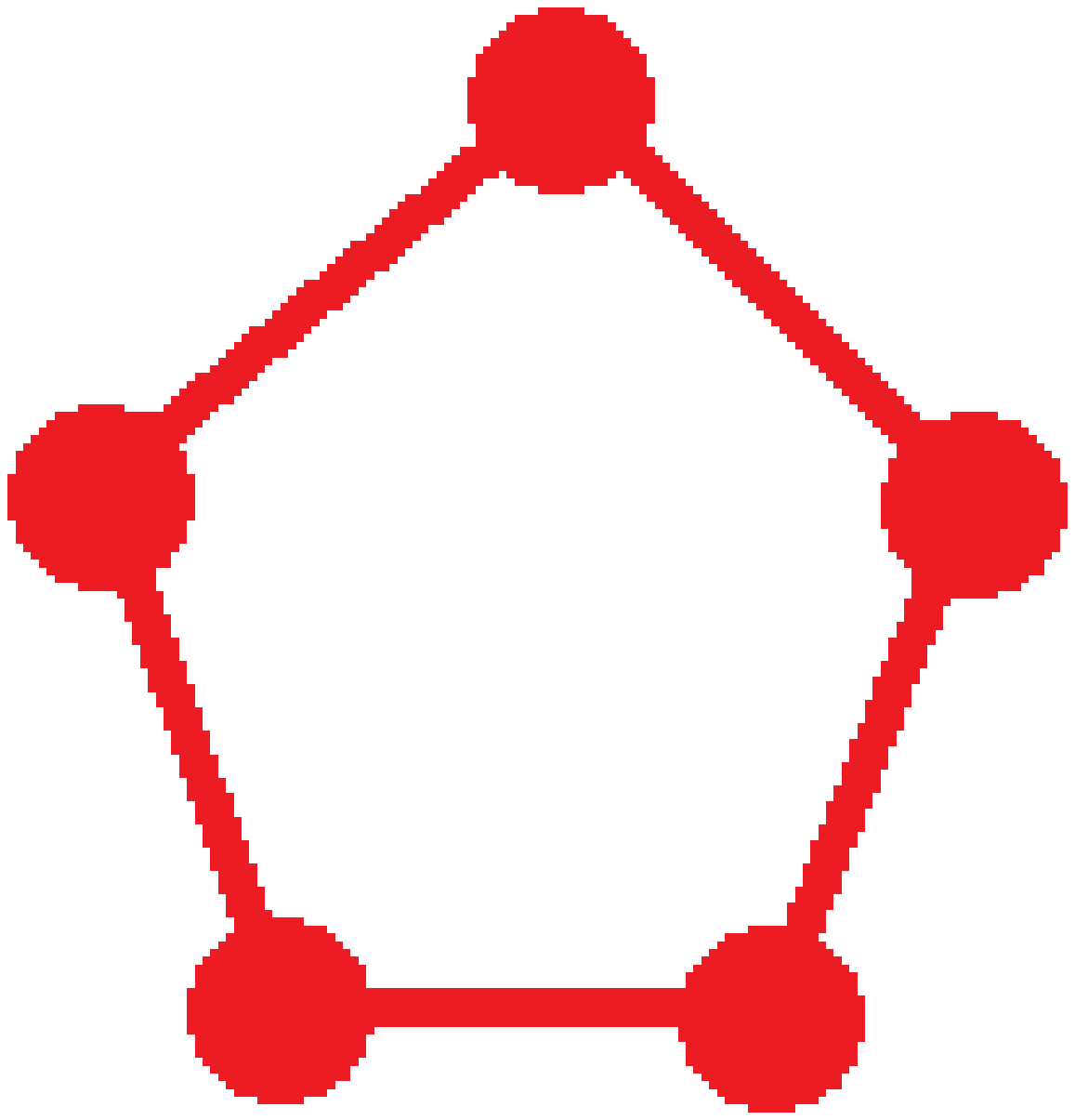}
\end{minipage}
\begin{minipage}{0.22\linewidth}
\centering
\includegraphics[width=0.6\linewidth]{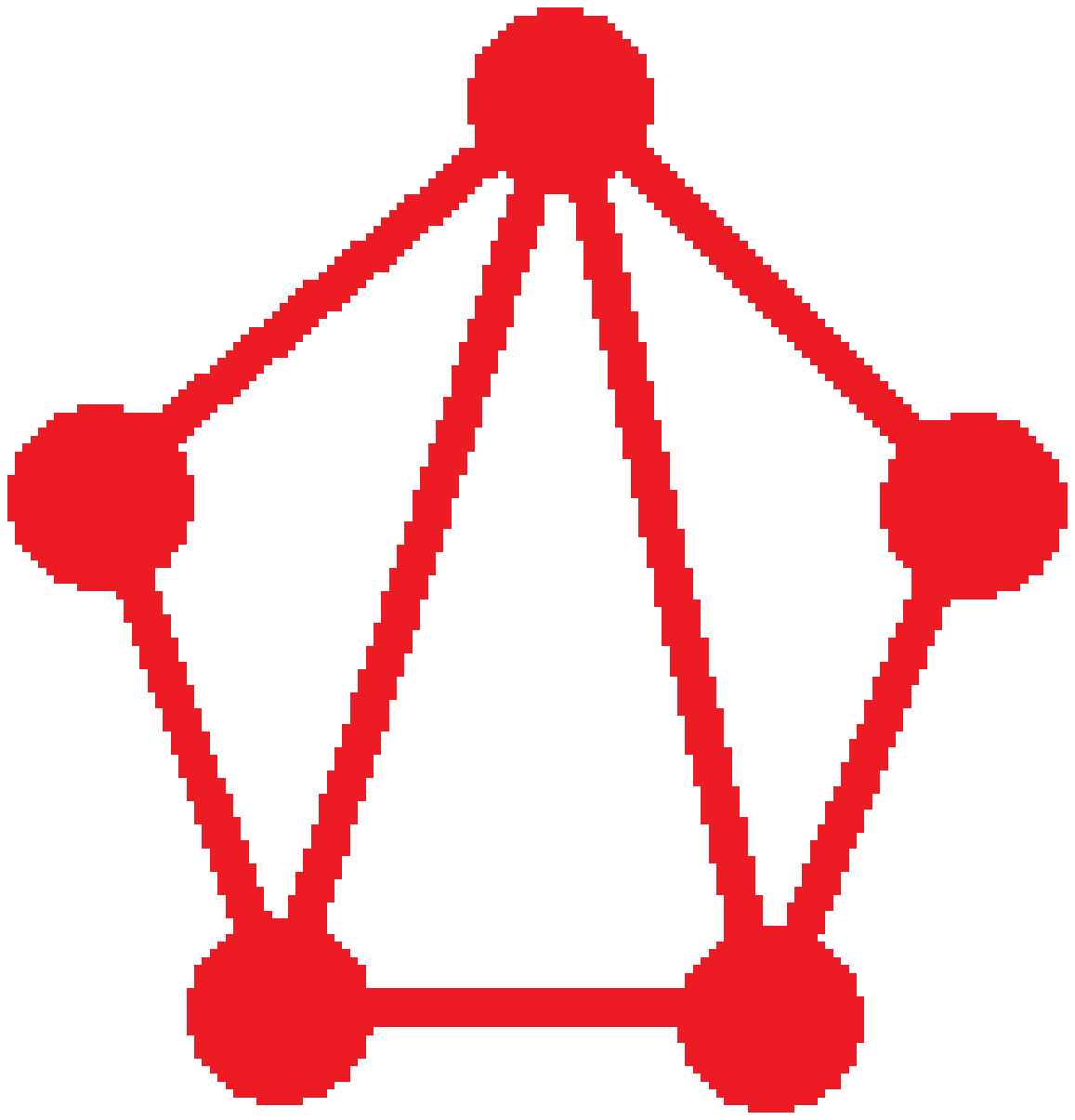}
\end{minipage}
\begin{minipage}{0.22\linewidth}
\centering
\includegraphics[width=0.6\linewidth]{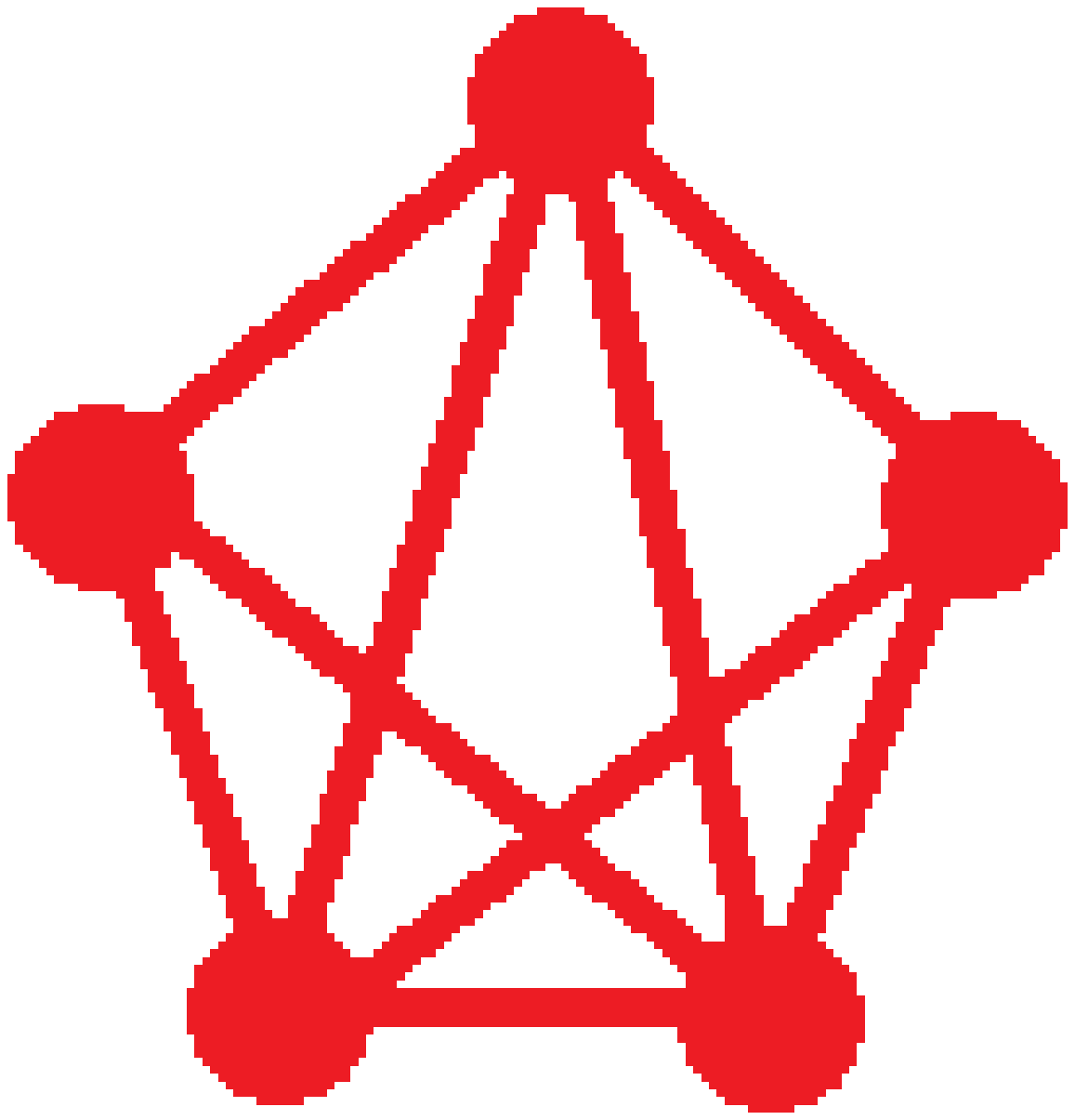}
\end{minipage}
\begin{minipage}{0.22\linewidth}
\centering
\includegraphics[width=0.6\linewidth]{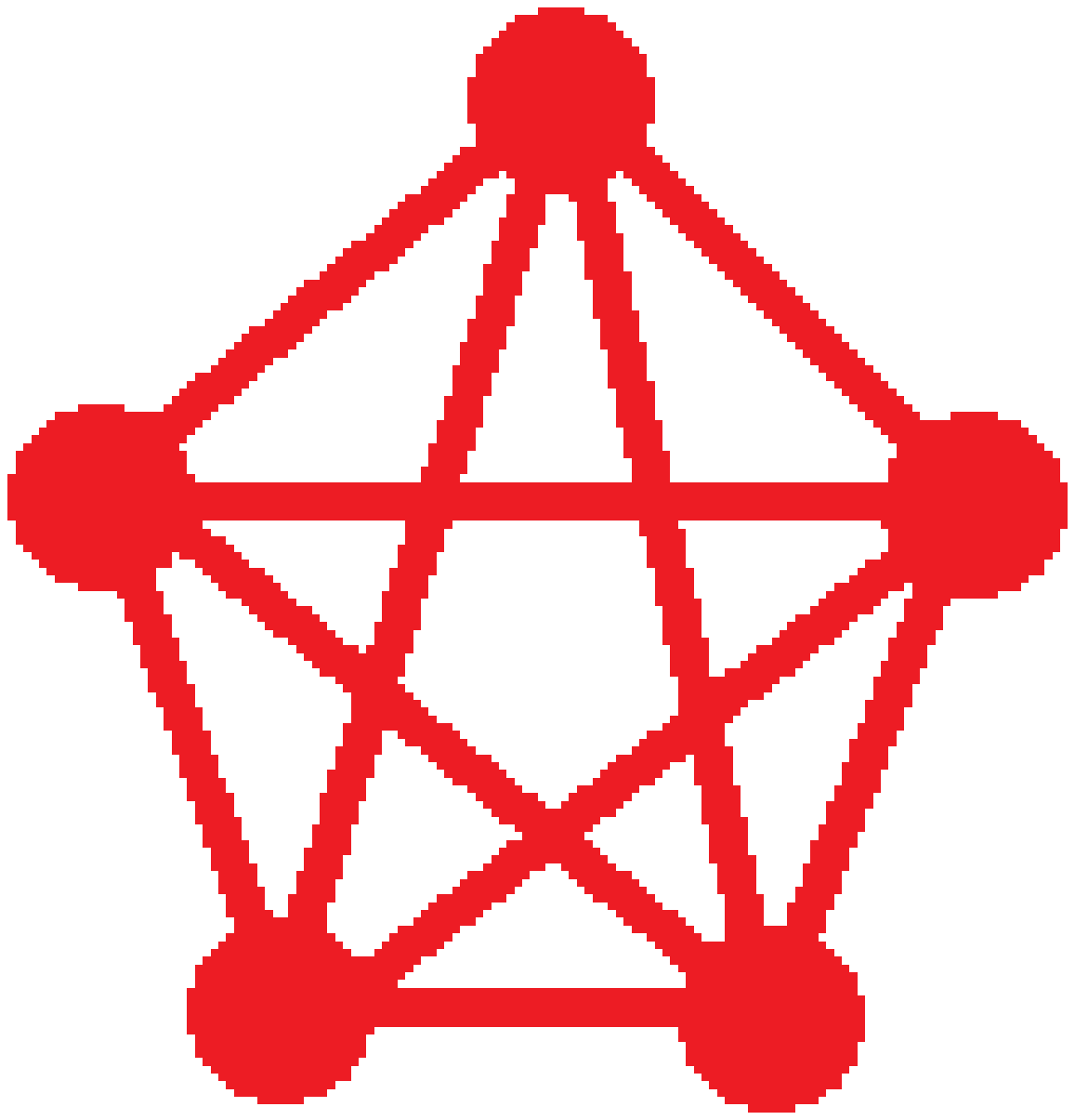}
\end{minipage}

\smallSpace
\begin{minipage}{0.22\linewidth}
\centering
(a) 0
\end{minipage}
\begin{minipage}{0.22\linewidth}
\centering
(b) 0.7
\end{minipage}
\begin{minipage}{0.22\linewidth}
\centering
(c) 0.9
\end{minipage}
\begin{minipage}{0.22\linewidth}
\centering
(d) \textbf{1}
\end{minipage}

\begin{minipage}{1\linewidth}
\textbf{\small Property 1: Internal Structure: } 
\end{minipage}	

\begin{minipage}{0.48\linewidth}
\centering
\includegraphics[width=0.5\linewidth]{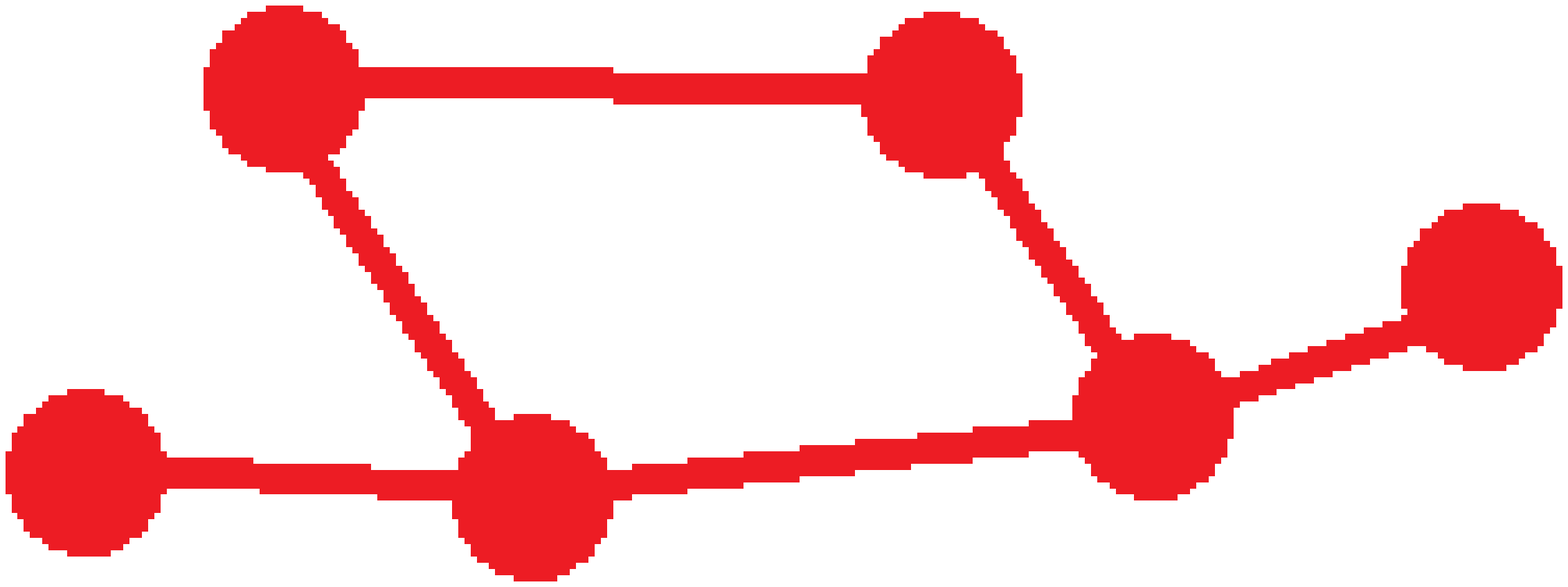}
\end{minipage}
\begin{minipage}{0.48\linewidth}
\centering
\includegraphics[width=0.5\linewidth]{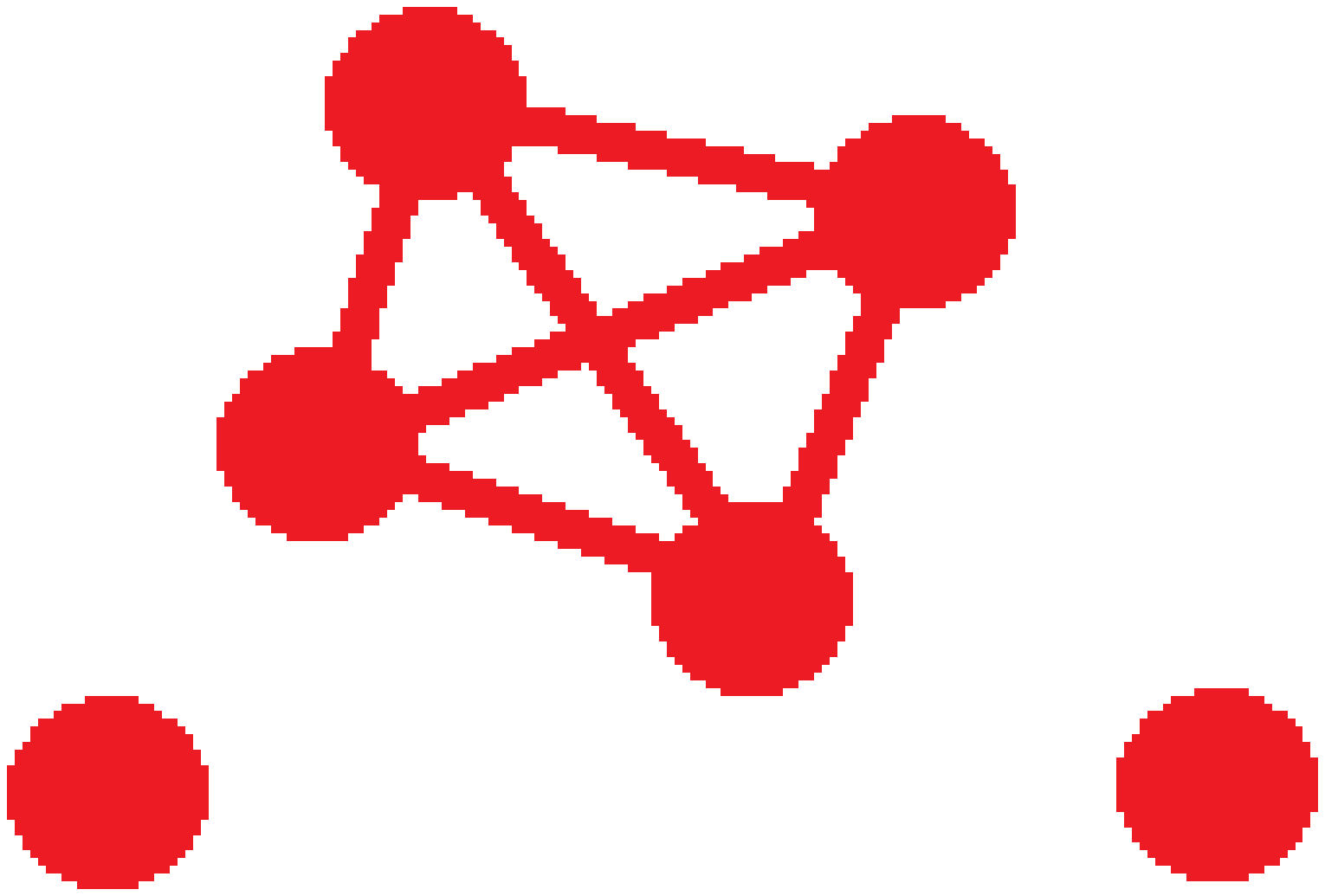}
\end{minipage}

\smallSpace
\begin{minipage}{0.48\linewidth}
\centering
(e) 0
\end{minipage}
\begin{minipage}{0.48\linewidth}
\centering
(f) \textbf{0.667}
\end{minipage}

\begin{minipage}{1\linewidth}
\textbf{\small Property 2: Linear Community Cohesion: }
\end{minipage}	

\begin{minipage}{0.22\linewidth}
\centering
\includegraphics[width=1\linewidth]{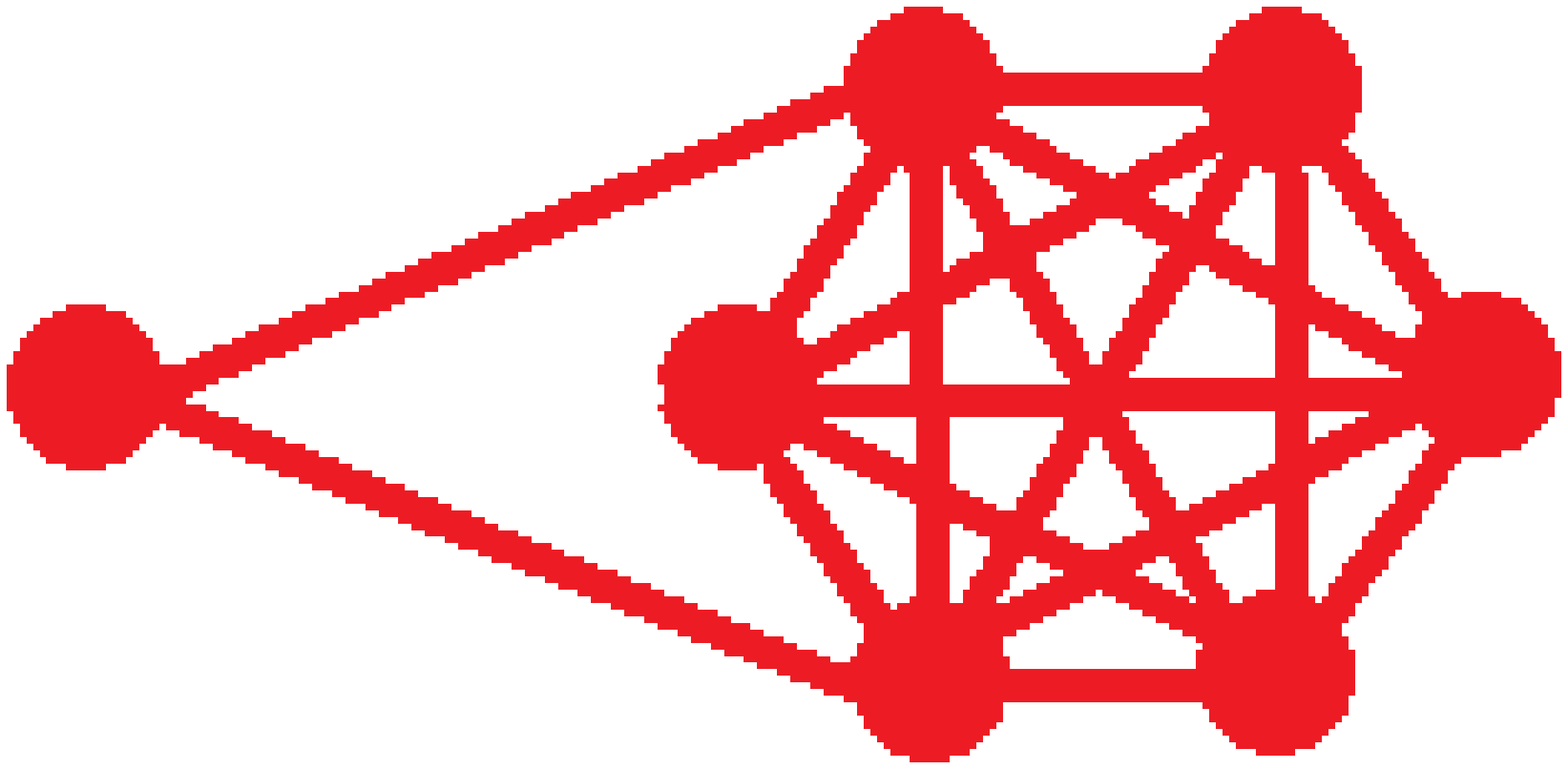}
\end{minipage}
\begin{minipage}{0.22\linewidth}
\centering
\includegraphics[width=1\linewidth]{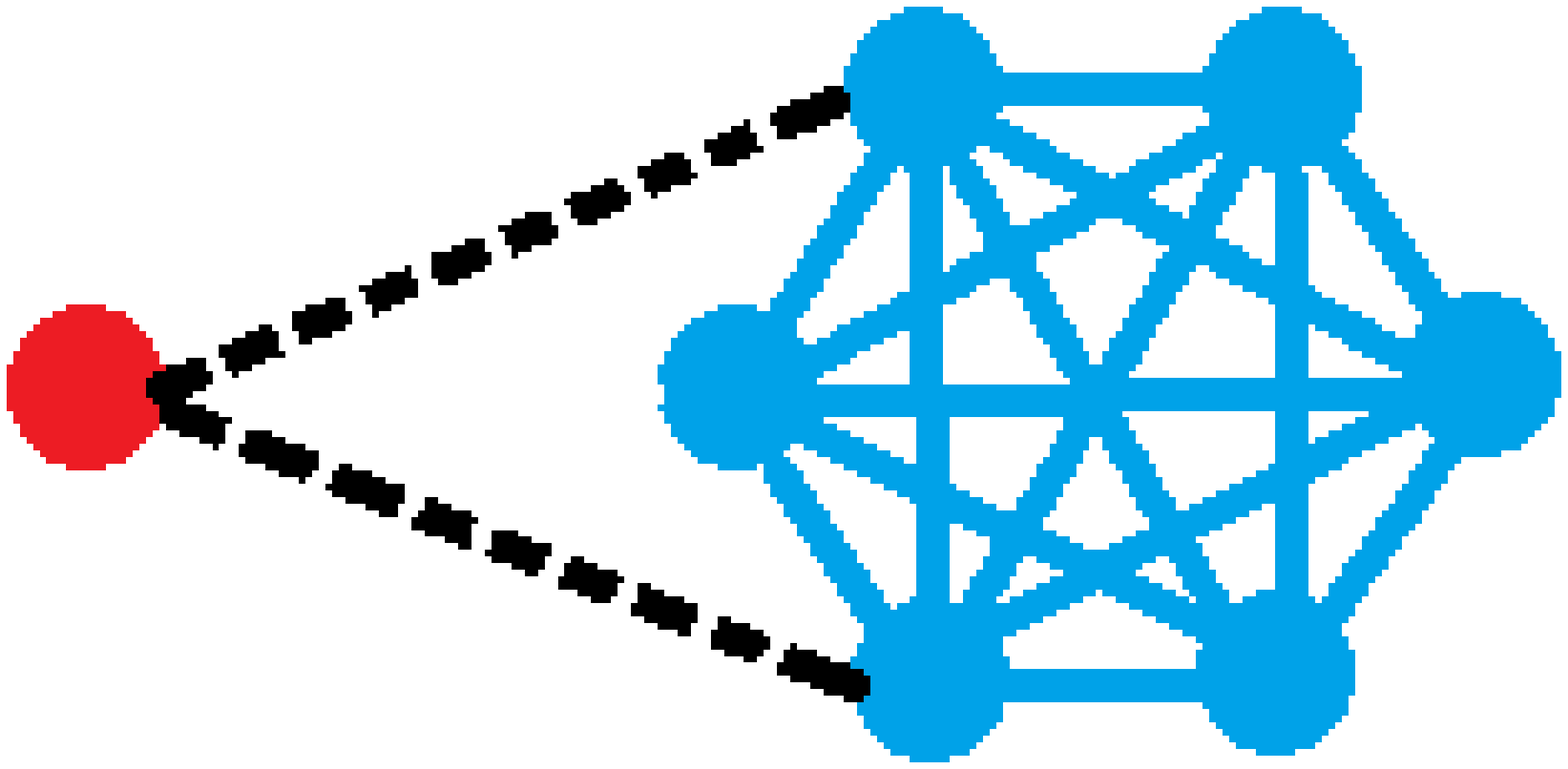}
\end{minipage}
\begin{minipage}{0.22\linewidth}
\centering
\includegraphics[width=1\linewidth]{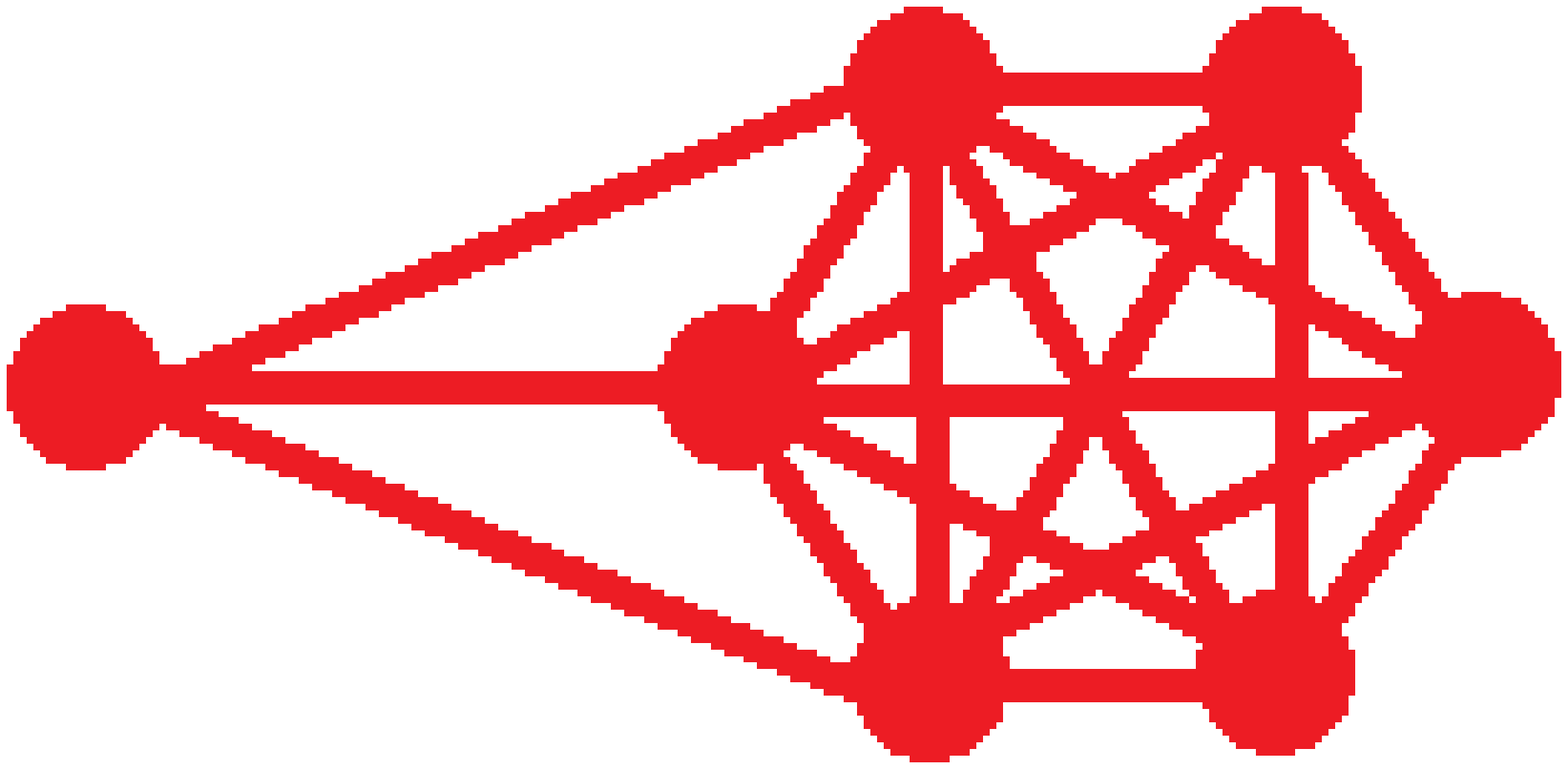}
\end{minipage}
\begin{minipage}{0.22\linewidth}
\centering
\includegraphics[width=1\linewidth]{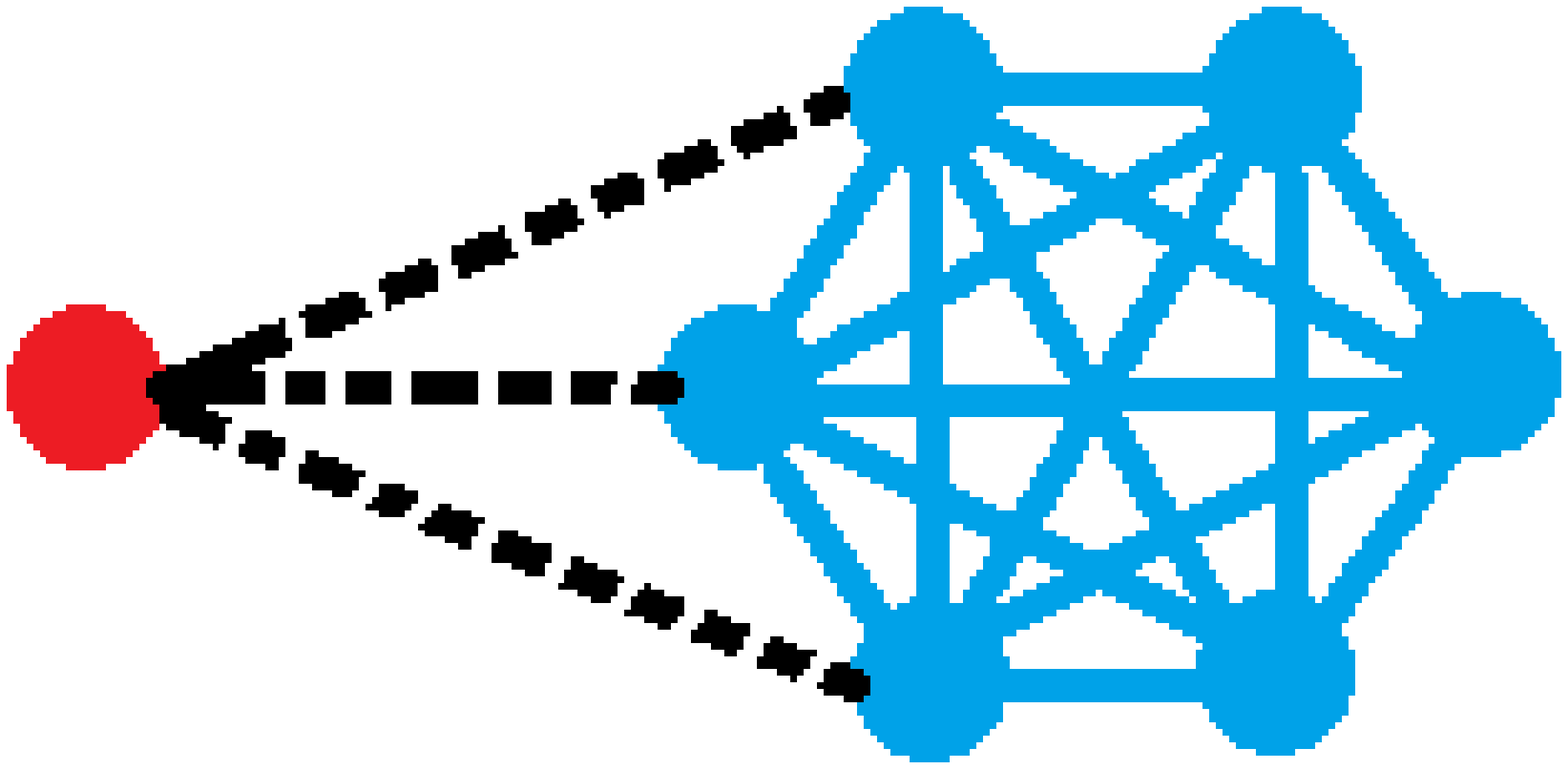}
\end{minipage}

\smallSpace
\begin{minipage}{0.22\linewidth}
\centering
(g) 0.833
\end{minipage}
\begin{minipage}{0.22\linewidth}
\centering
(h) \textbf{0.860}
\end{minipage}
\begin{minipage}{0.22\linewidth}
\centering
(i) \textbf{0.881}
\end{minipage}
\begin{minipage}{0.22\linewidth}
\centering
(j) 0.814
\end{minipage}

\begin{minipage}{1\linewidth}
\textbf{\small Property 3: Bridges:}
\end{minipage}

\begin{minipage}{0.45\linewidth}
\centering
\includegraphics[width=0.7\linewidth]{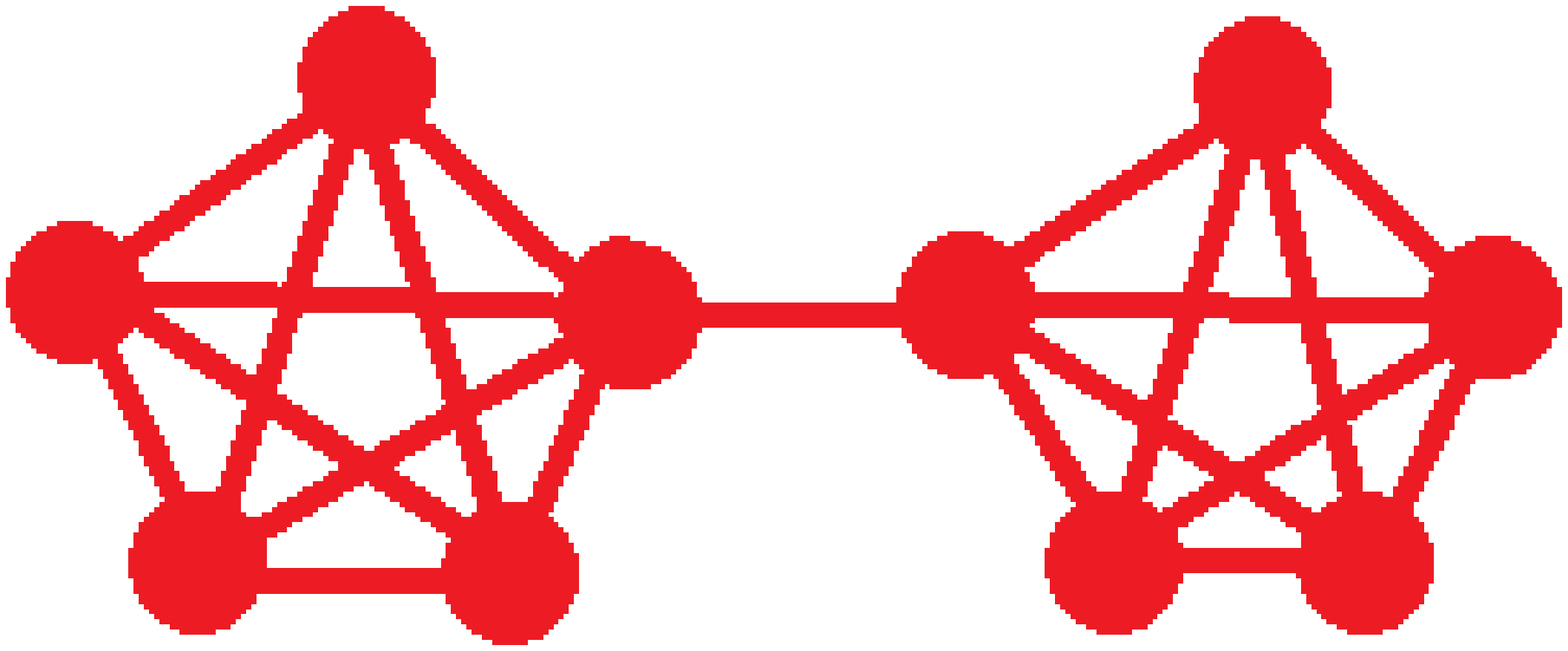}
\end{minipage}
\begin{minipage}{0.45\linewidth}
\centering
\includegraphics[width=0.7\linewidth]{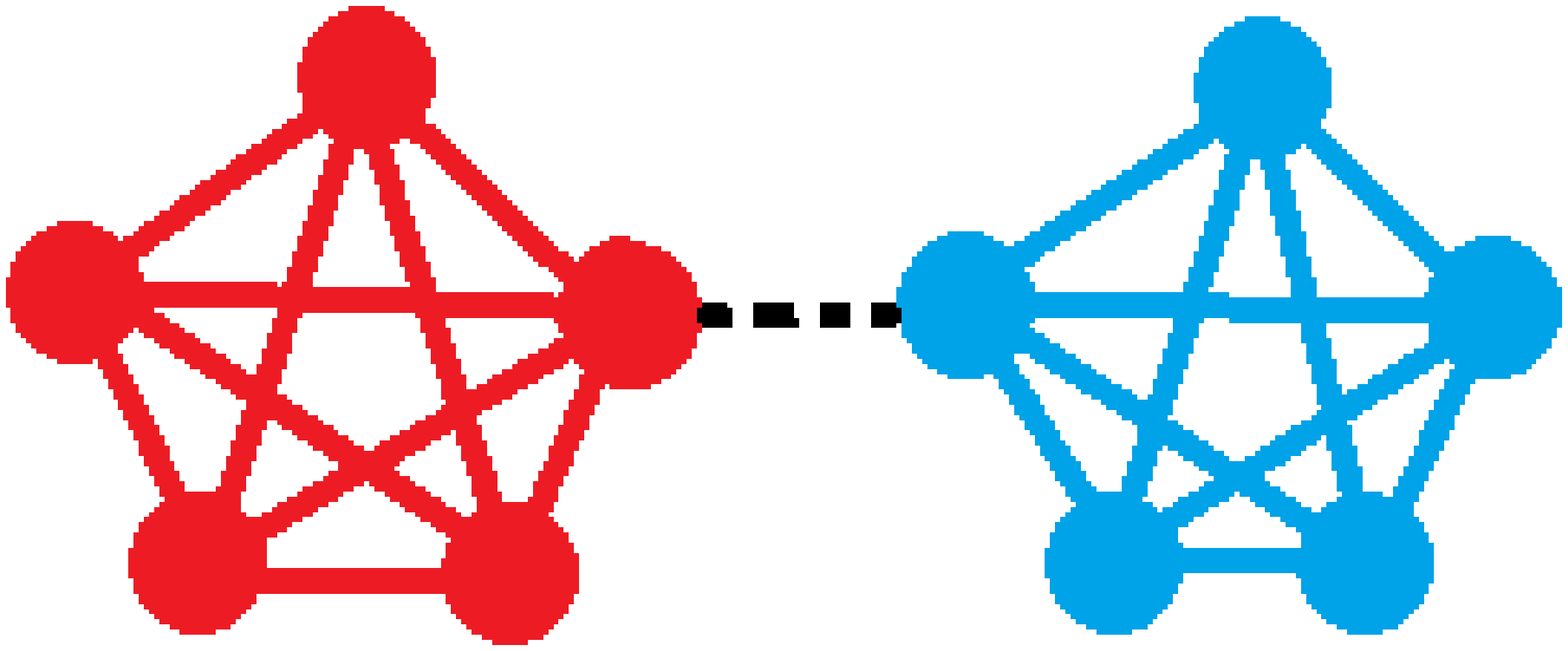}
\end{minipage}

\smallSpace
\begin{minipage}{0.45\linewidth}
\centering
(k) 0.444
\end{minipage}
\begin{minipage}{0.45\linewidth}
\centering
(l) \textbf{1}
\end{minipage}

\begin{minipage}{1\linewidth}
\textbf{\small Property 4: Cut Vertex Density:}
\end{minipage}
\begin{minipage}{0.32\linewidth}
\centering
\includegraphics[width=1\linewidth]{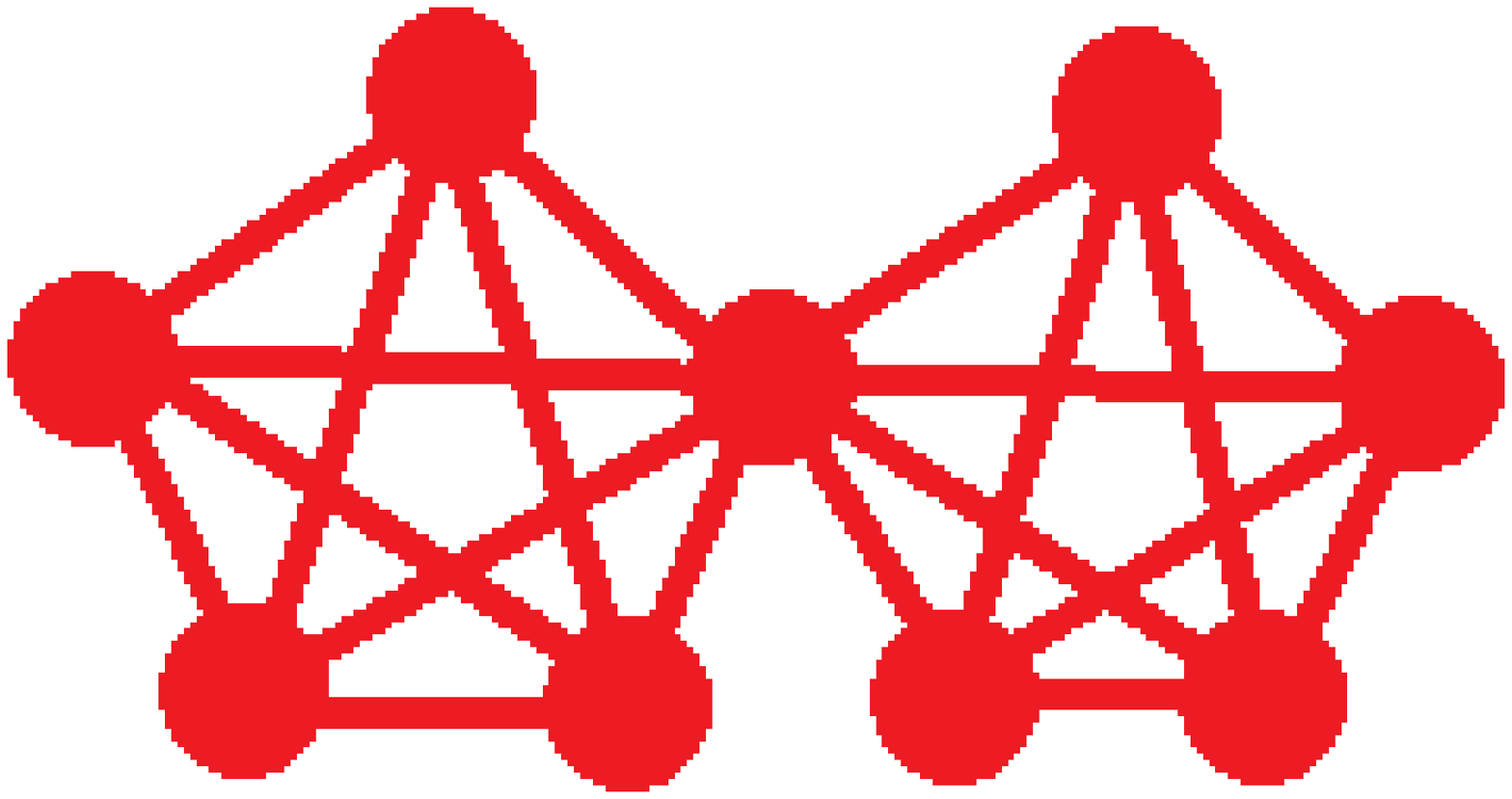}
\end{minipage}
\begin{minipage}{0.32\linewidth}
\centering
\includegraphics[width=1\linewidth]{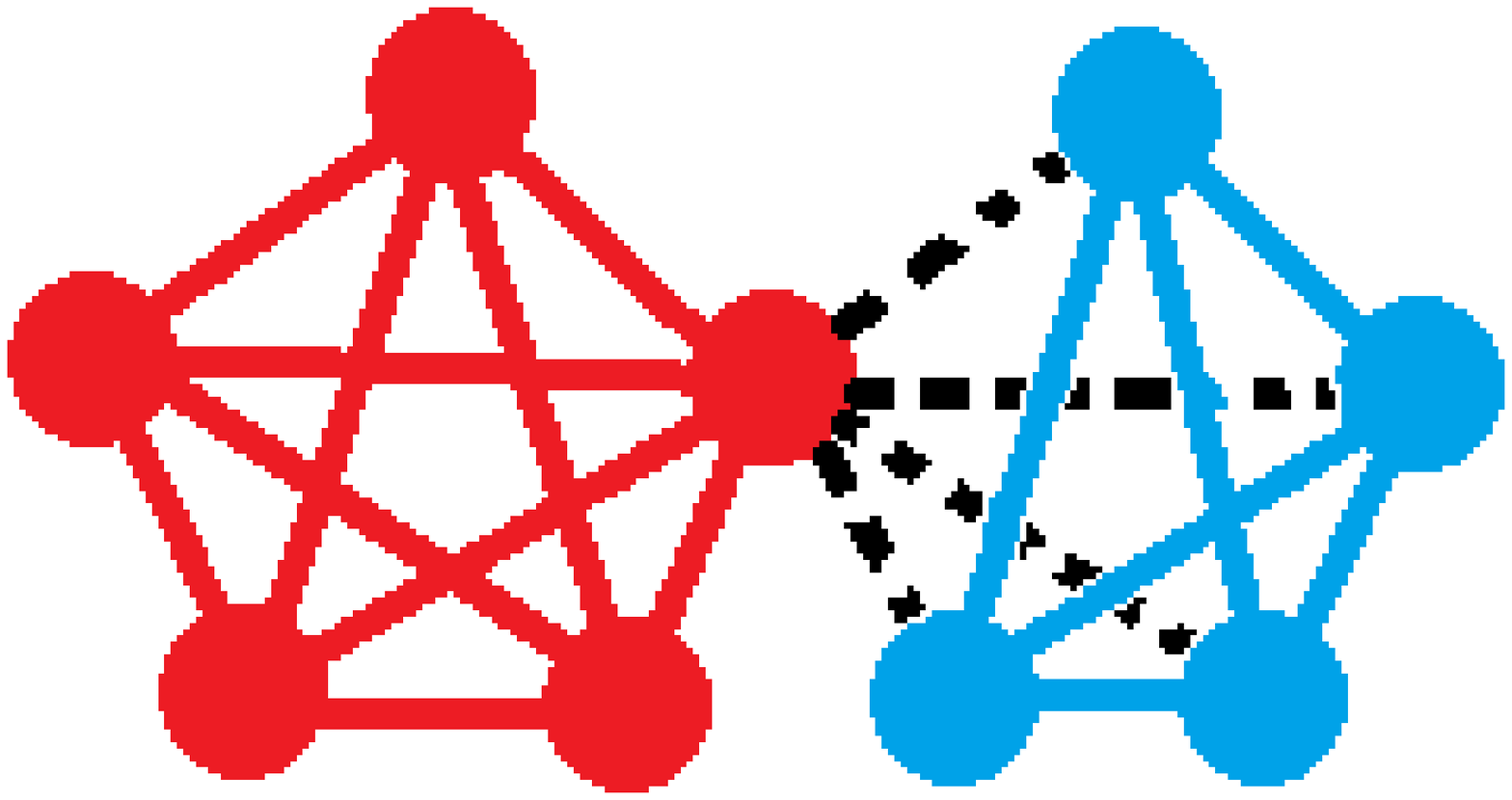}
\end{minipage}
\begin{minipage}{0.32\linewidth}
\centering
\includegraphics[width=1\linewidth]{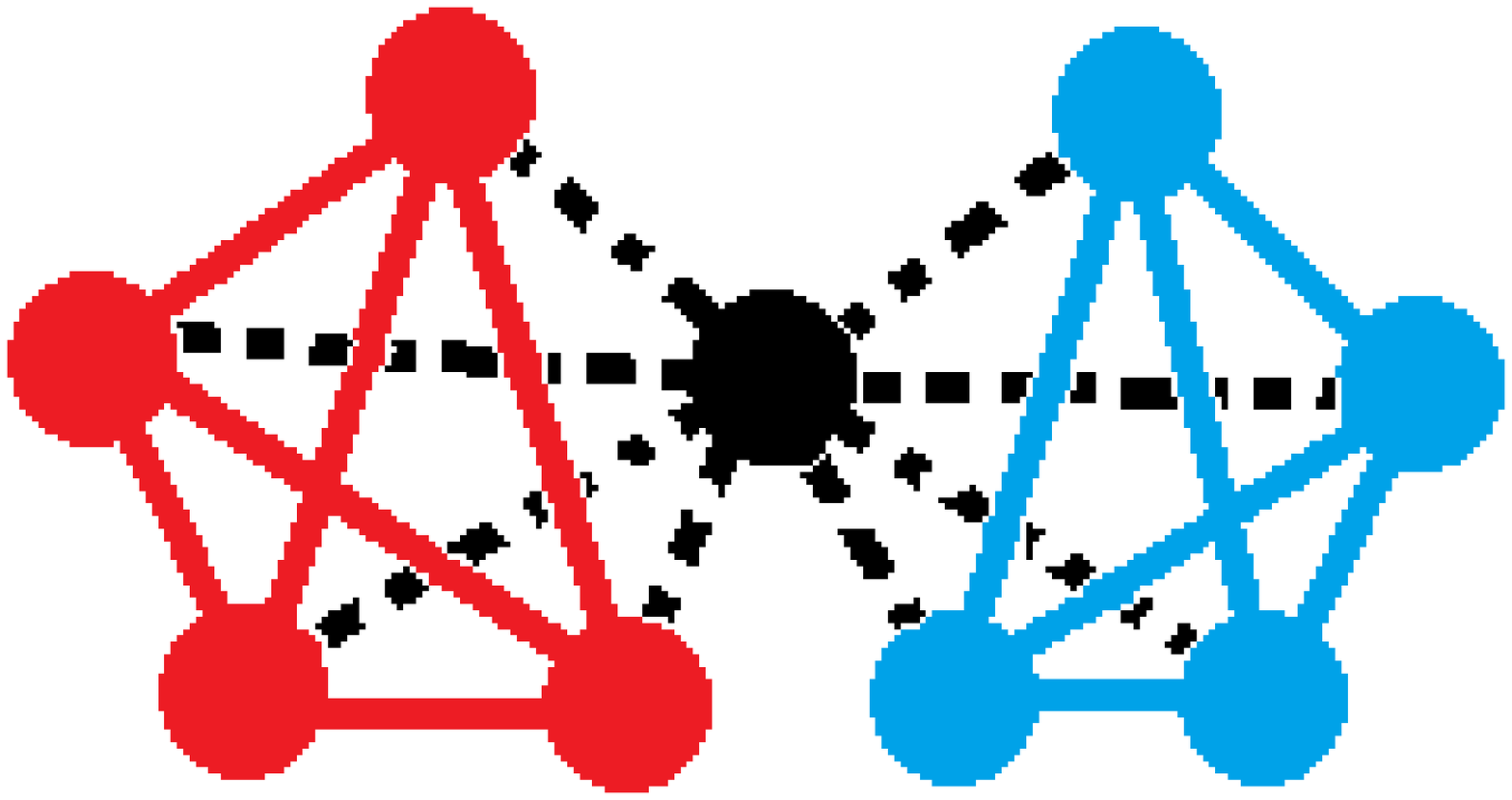}
\end{minipage}

\smallSpace
\begin{minipage}{0.32\linewidth}
\centering
(m) 0.556
\end{minipage}
\begin{minipage}{0.32\linewidth}
\centering
(n) \textbf{0.722}
\end{minipage}
\begin{minipage}{0.32\linewidth}
\centering
(o) 0.444
\end{minipage}

\caption{Property examples.}
\label{fig:exemple_casos_teoremes}
\end{figure}

The clique is the subgraph structure that best resembles the 
perfect community, and thus, 
$WCC$ rates it with the largest value. 
On the other hand, if the community has no triangles, 
its quality is the minimum possible. 
In Figure~\ref{fig:exemple_casos_teoremes}(a-d), we show a 
community of five vertices with an increasing
number of internal triangles. The larger the triangles density,
the larger the $WCC(S)$ value of the community.

\subsection{Properties}\label{section:behavior}

In this section, we introduce a set of basic properties that any community cohesion 
metric for social networks should fulfill. We verify them for $WCC$, 
proving that $WCC$ is a good candidate to distinguish
communities in social networks (proofs are given in
Appendices~\ref{appendix:primer_teorema}-\ref{appendix:formal_proof_end}):.

{\flushleft \textbf{Property 1: Internal Structure.}}
In several previous studies~\cite{boguna2004models,newman2003social}, it
has been proved that one of the main characteristics of social networks 
is the presence of a large clustering coefficient and communities. Social networks have more triangles than
expected in random graphs~\cite{newman2003social,newman2001structure,
shi2007networks,satuluri2011local} and models describing the growth of social networks give 
triangle closing as a key factor of network evolution~\cite{leskovec2008microscopic}.
Thus, we take triangles as the indicator of the presence of community structures. Then, \textit{\textbf{the cohesion of a community
given by a community metric for social networks, depends on the triangles formed by the edges inside the community.}}
We verify this property for $WCC$: the left factor in Equation~(\ref{eq:ccc_normalizat2}) is the ratio of the number of triangles the vertex $x$
forms with the vertices in $S$ as opposed to the number of triangles the vertex $x$ forms with the whole graph. Hence, the factor is affected by the number
of triangles inside the community. On the other hand, the right factor depends on the number of vertices that form triangles with vertex $x$. Therefore, the
distribution of the triangles inside the community affect the right factor.
Figure~\ref{fig:exemple_casos_teoremes}(e-f) shows an example of two partitions with the same number of edges, but distributed differently.
We see that the vertices in Figure~\ref{fig:exemple_casos_teoremes}(e) form no triangles which
translates to a value of $WCC=0$. On the other hand, the vertices in Figure~\ref{fig:exemple_casos_teoremes}(f) form four triangles,
obtaining a larger value of $WCC$. We see that $WCC$ reacts to the internal structure of the communities, and in particular to the
presence of triangles. 

{\flushleft \textbf{Property 2: Linear Community Cohesion.}}
An interesting 
aspect to consider is the dynamics of community formation in social networks:
what happens when there is an existent community and a new vertex appears in 
a graph, which is creating links with the members of the community. 
In order to keep high quality communities, these must grow with cohesion. 
This means that a vertex can
only join a community if it has a significant 
number of links with the members of the community.
The larger the community, the more links are needed. Otherwise,
the cohesion of the community decreases. This simple
restriction limits the community growth if there is not a significant cohesion
among its members. Therefore, 
\textit{\textbf{the number of connections needed between a vertex $x$ and a set $S$, so that  
$f(S\cup\{x\}) \geq f(S,\{x\})$, grows linearly with respect to the size of $S$}}. 
If it grew sublinearly, it would mean that the larger a community is, the easier would be 
for a vertex to join the community relative to the community size.
On the other hand, if it grew faster than linear, the communities would have a maximum possible size.

Theorem~\ref{theorem:vertex_contra_comunitat} proves 
this requirement for $WCC$: 
\begin{teorema}
\teoremaCliqueNode
\label{theorem:vertex_contra_comunitat}
\end{teorema}
For instance, in the particular case of the clique (where $p=1$), it is necessary
to connect to roughly more than one third of the vertices to become a member
of the community.
\begin{corolari}
\corolariClique
\label{corollary:clique}
\end{corolari}
In Figure~\ref{fig:exemple_casos_teoremes}(g-j) we show an example of
Theorem~\ref{theorem:vertex_contra_comunitat}, where colors indicate different
communities, and dashed lines represent edges between communities. 
In (g) and (i), the whole graph is a community but
in (h) and (j) the graph is split into two communities. 
When the external vertex
has only two connections with the six vertices, the metric considers
better to keep the vertex outside of the community. 
However, when the number of connections
is three, $WCC$ has a better value when the vertex is 
included into the community.

{\flushleft \textbf{Property 3: Bridges.}} 
A bridge is an edge that
if it is removed from the graph, it creates two connected components. 
The connections in real graphs are known not to be local, but can connect
distant vertices~\cite{liben2007link}. A bridge is a very weak relation between
two sets of vertices that are unrelated, because it only affects one
member of both datasets. 
Therefore, 
\textit{\textbf{an optimal community in social networks can not contain a bridge.}}
We prove that $WCC$ is resistent to bridges in the 
following theorem:

\begin{teorema}
\teoremaBridgeEdge
\label{theorem:bridge_edge}
\end{teorema}



An edge that does not close any triangle, does not affect 
the computation of $WCC$ because it alters no terms in 
Equation~(\ref{eq:ccc_normalizat2}). 
A bridge is a particular case of such an edge, and therefore,
it does not affect the quality of a partition for $WCC$. 
Since given two connected components it is better
to have them separated than merged into a single community, then 
an optimal community cannot contain a 
bridge, because there exists a partition with a better cohesion 
formed by the two separated components.
In Figure~\ref{fig:exemple_casos_teoremes}(k-l) we show an example of the application of 
Theorem~\ref{theorem:bridge_edge}. We see that having the 
two cliques separated is better than considering a single community with a bridge, in terms of $WCC$.

{\flushleft \textbf{Property 4: Cut Vertex Density.}} 
A cut vertex is a vertex whose removal disconnects the graph into two or more
connected components.
A cut vertex is certainly a weak link 
in a community formed by the union of the
two sets, because the vertices of the two sets have no relation among them.
However, if the two sets have no other connection among them 
rather than the cut vertex, the two sets must be considered 
as independent communities on 
their own if they have enough cohesion internally.  
Therefore, 
\textit{\textbf{an optimal community can not contain a cut vertex if the sets that
it separates have a minimum density}}\footnote{A vertex cut 
can be seen as an example of an overlapped community. However,
it is not the aim of this work to consider 
the problem of overlapping communities.}.
In Figure~\ref{fig:exemple_casos_teoremes}(m-o), we show 
two cliques (note that the clique is the highest density graph structure)
of size five sharing a vertex. Here, 
$WCC$ is able
to separate the communities for this particular case because the red and blue
sets of vertices have enough cohesion to become separate communities.
We prove this property for $WCC$ for the case where communities have the highest
possible density, which is the clique:

\begin{teorema}
\teoremaCliqueClique
\label{theorem:cliques_no_es_junten}
\end{teorema}

This theorem illustrates the fact $WCC$ avoids merging two very well 
defined communities (such as two cliques)
because of a single vertex. The reason is that $WCC$ is a metric 
that not only takes into account 
the vertices that are connected and form triangles, but also the vertices 
that do not. Thus, if the triangles inside the community are not 
distributed evenly among all the vertices, then the quality of
the community is penalized.

\subsection{Examples of $WCC$ on communities}

\begin{figure}[t!]
\begin{minipage}{\gridsize\linewidth}
\centering
\includegraphics[scale=0.15]{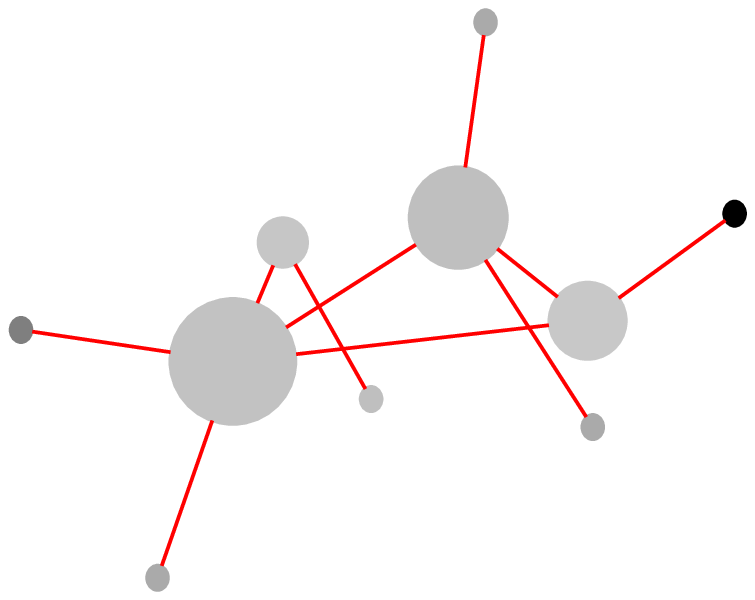}
\end{minipage}
\begin{minipage}{\gridsize\linewidth}
\centering
\includegraphics[scale=0.15]{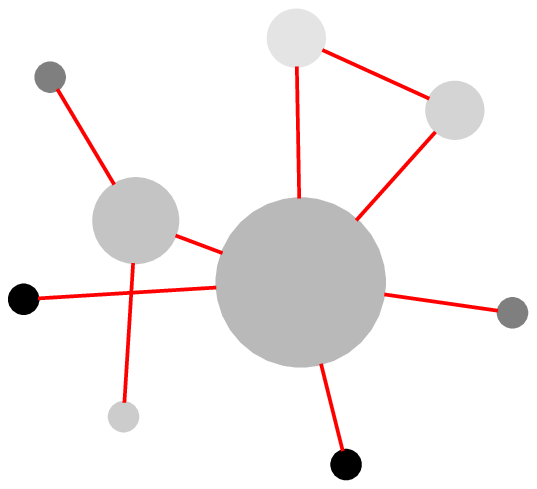}
\end{minipage}
\begin{minipage}{\gridsize\linewidth}
\centering
\includegraphics[scale=0.15]{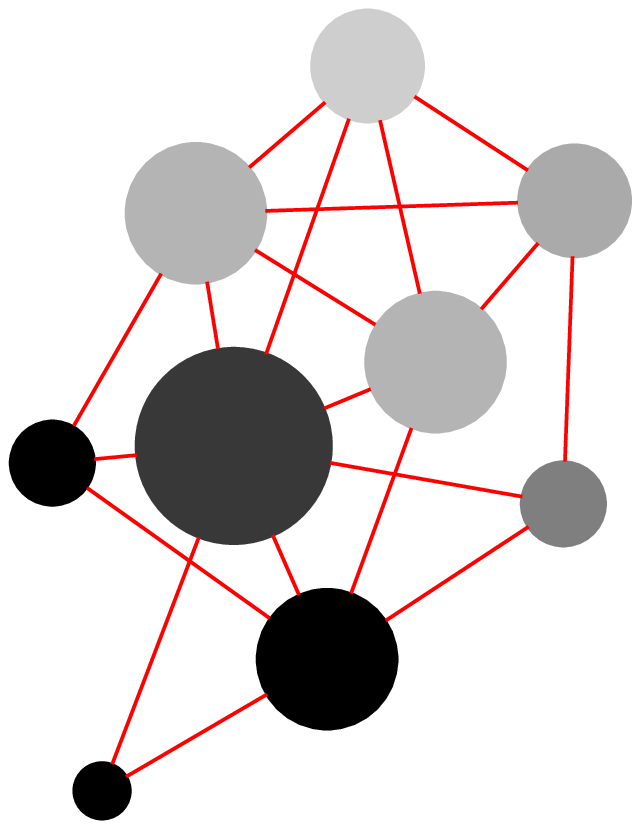}
\end{minipage}
\begin{minipage}{\gridsize\linewidth}
\centering
\includegraphics[scale=0.15]{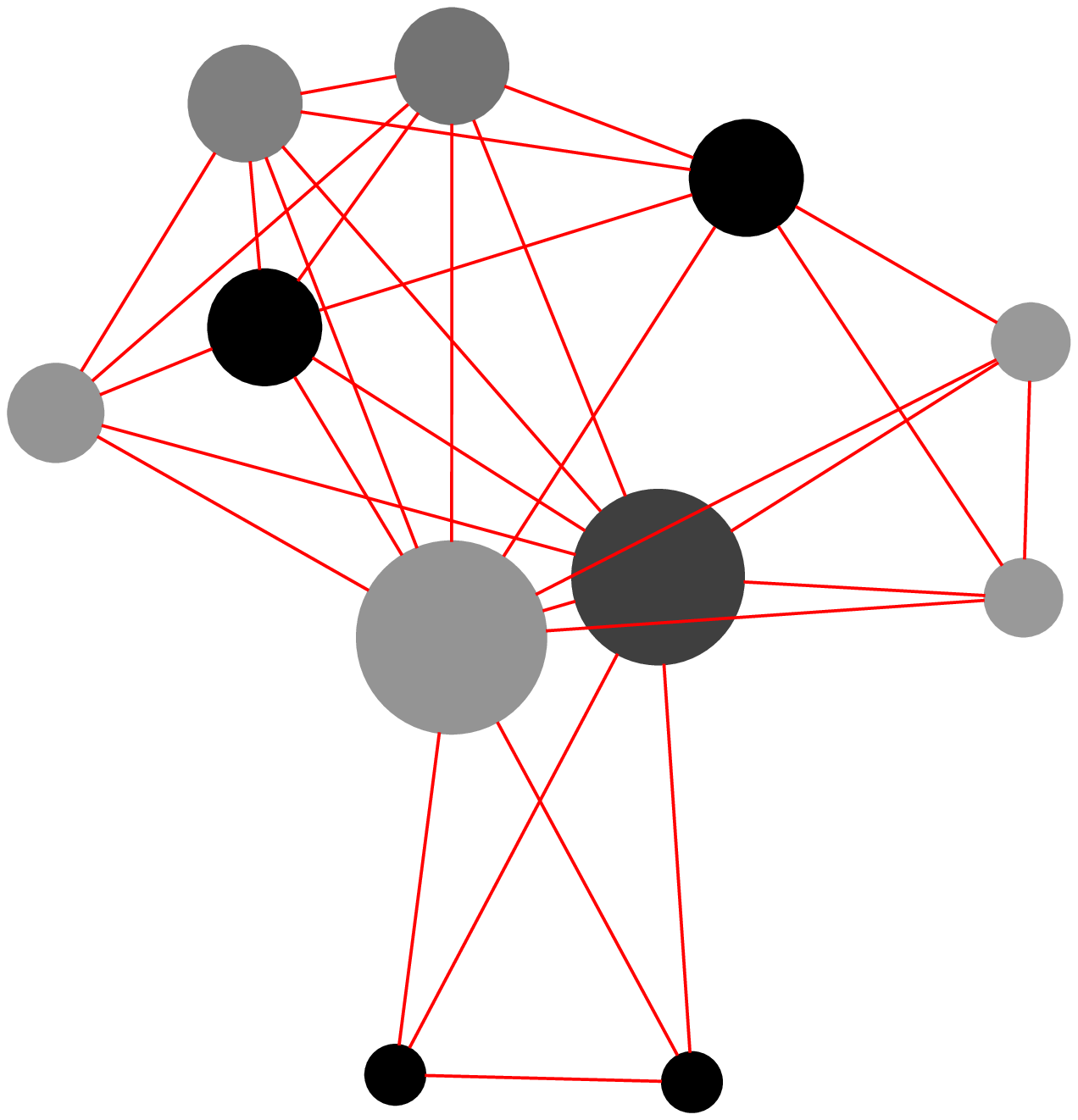}
\end{minipage}

\begin{minipage}{\gridsize\linewidth}
\centering
(a) 0.09048
\end{minipage}
\begin{minipage}{\gridsize\linewidth}
\centering
(b) 0.14191
\end{minipage}
\begin{minipage}{\gridsize\linewidth}
\centering
(c) 0.25457
\end{minipage}
\begin{minipage}{\gridsize\linewidth}
\centering
(d) 0.31792
\end{minipage}

\begin{minipage}{\gridsize\linewidth}
\centering
\includegraphics[scale=0.15]{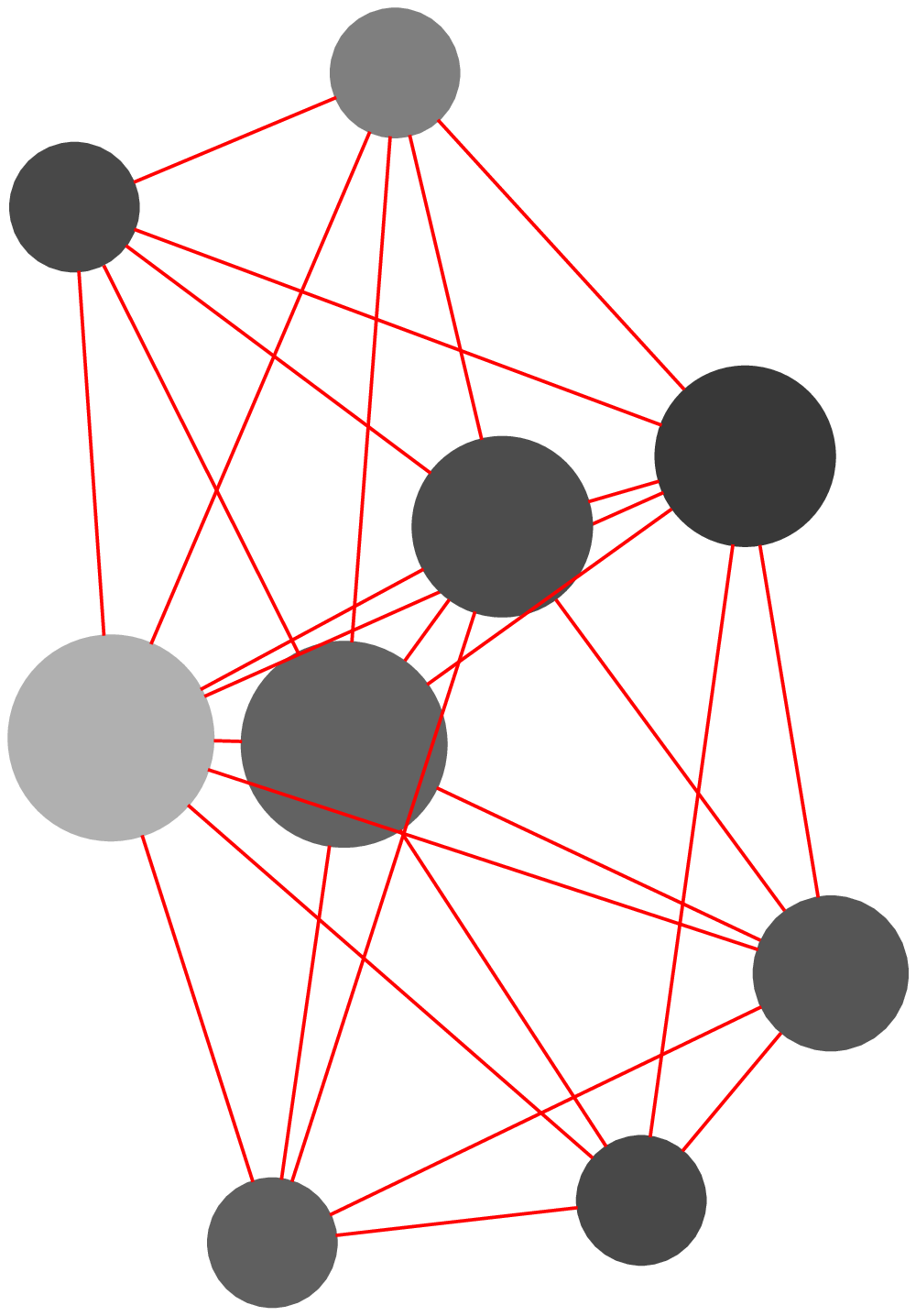}
\end{minipage}
\begin{minipage}{\gridsize\linewidth}
\centering
\includegraphics[scale=0.15]{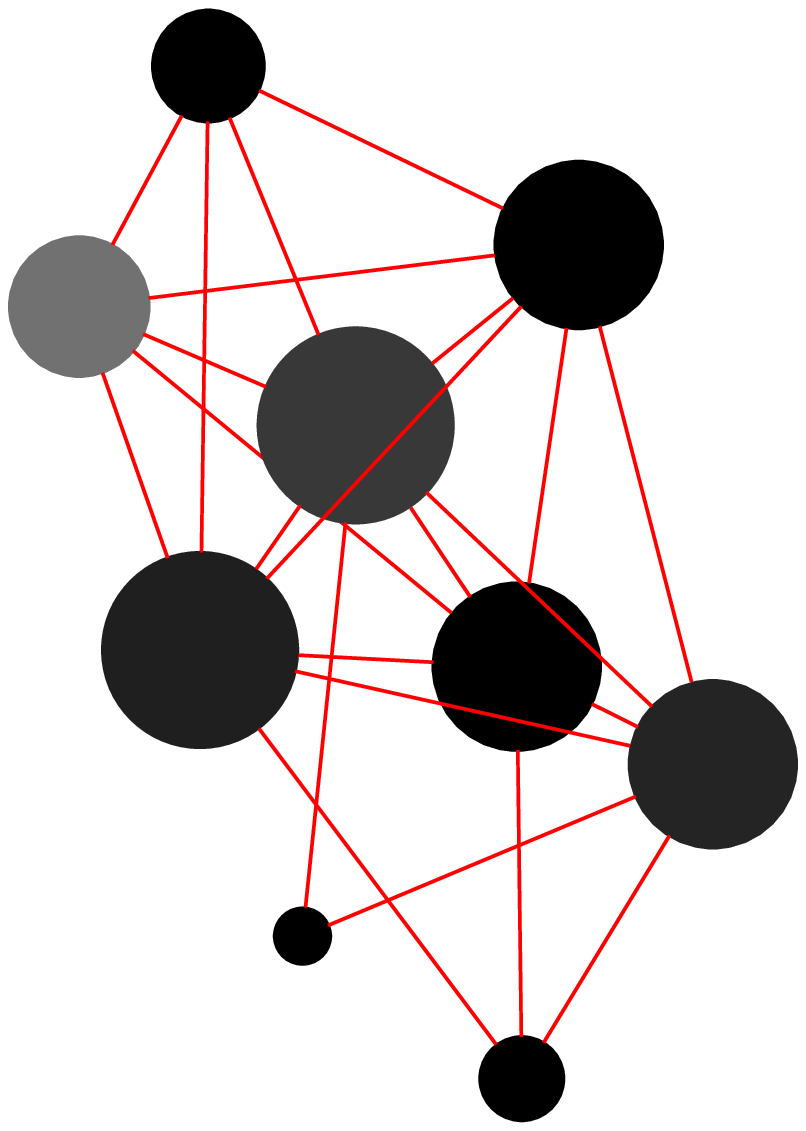}
\end{minipage}
\begin{minipage}{\gridsize\linewidth}
\centering
\includegraphics[scale=0.15]{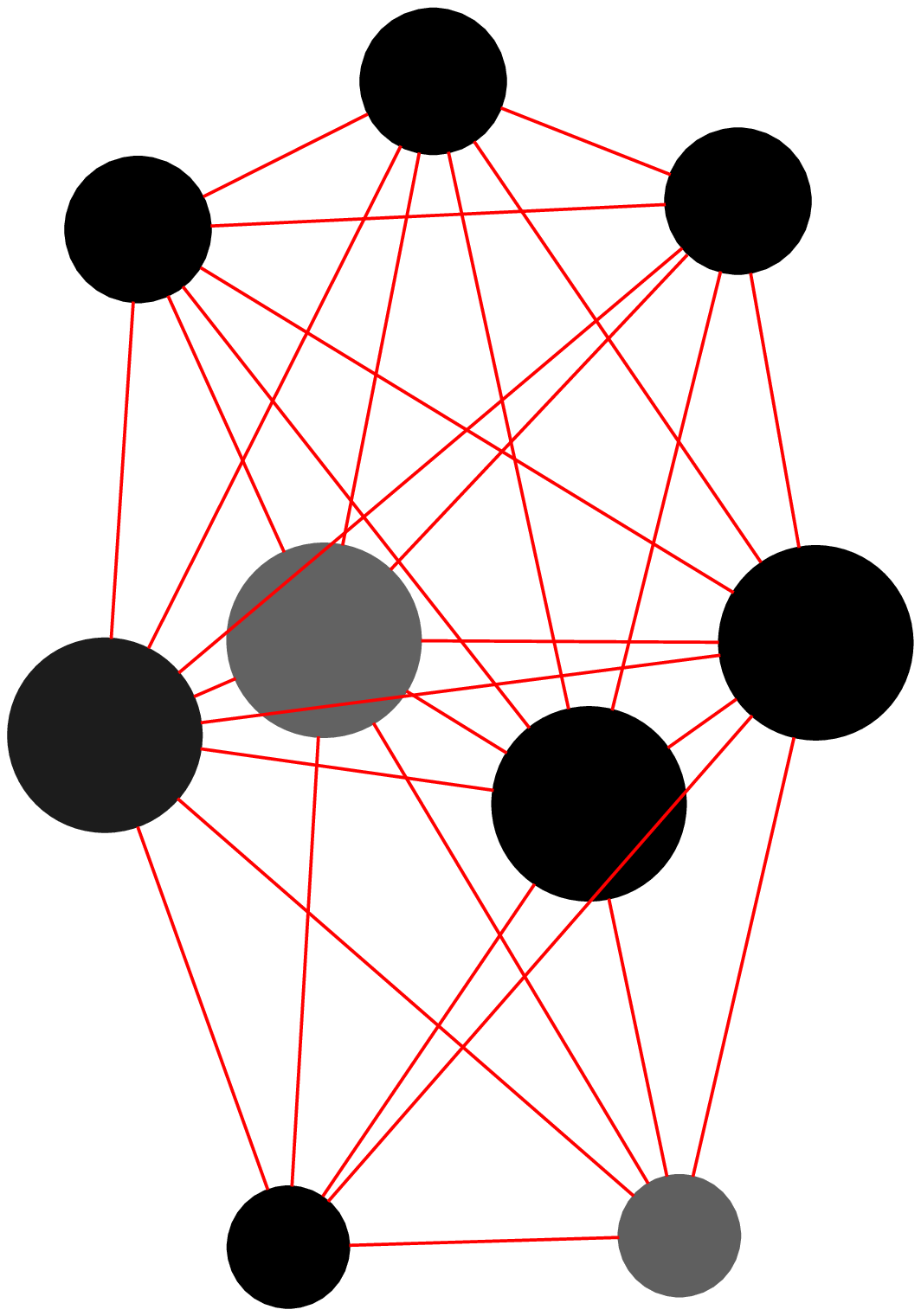}
\end{minipage}
\begin{minipage}{\gridsize\linewidth}
\centering
\includegraphics[scale=0.15]{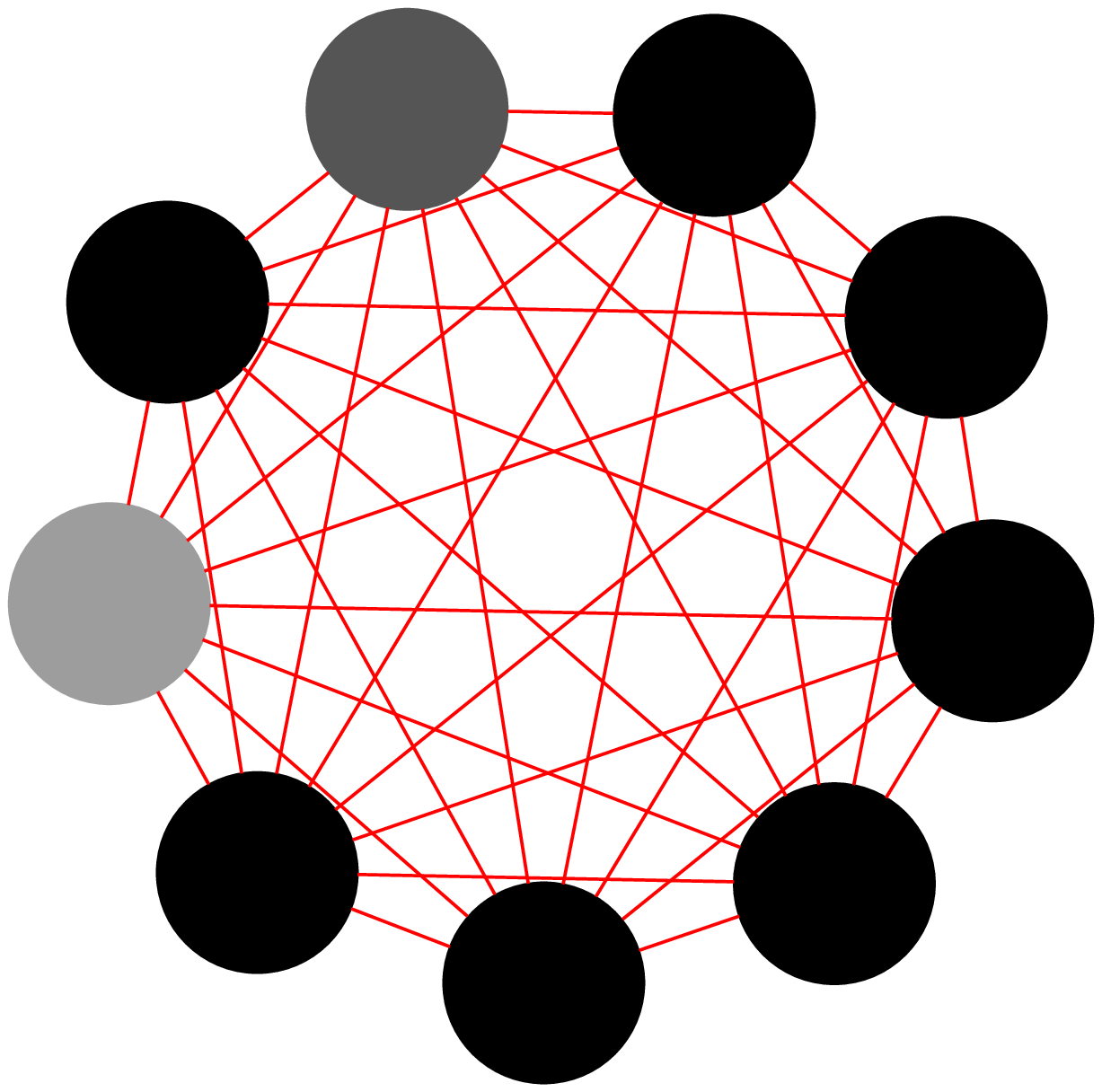}
\end{minipage}

\begin{minipage}{\gridsize\linewidth}
\centering
(e) 0.52118
\end{minipage}
\begin{minipage}{\gridsize\linewidth}
\centering
(f) 0.65128
\end{minipage}
\begin{minipage}{\gridsize\linewidth}
\centering
(g) 0.78072
\end{minipage}
\begin{minipage}{\gridsize\linewidth}
\centering
(h) 0.92798
\end{minipage}
\caption{Examples of communities from real graphs, sorted by
$WCC$.}
\label{fig:exemple_comunitats}
\end{figure}

Figure~\ref{fig:exemple_comunitats} shows some examples of communities with
different values of $WCC$. These communities are extracted randomly from the
set of communities found in the real graphs used by the algorithms in
Section~\ref{section:experiments}. The color of the vertices represents the
percentage of neighbors belonging to the community.
The darker the vertex, the
larger the percentage of neighbors of the vertex that belong to the community. On
the other hand, the size of the vertices represents the percentage of vertices of the
community that are actual neighbors of that vertex. The larger the size of the
vertex, the more connected the vertex is with the other vertices of the community. In
other words, the color represents the size of the edge cut that 
disconnects the vertex from the rest of the graph,
while the size represents the density of edges of
the vertex that connects it with other vertices in the community. Thus, the better
the community is, the larger and darker are its vertices. In the figure, we see
that the larger the $WCC$ of the community, the larger and darker are the
vertices of the community, which means that they are more densely connected and
better isolated from the rest of the graph. We see then, that there is a
correlation between high $WCC$ values and good communities.

\section{Comparison with other metrics}\label{section:comparison}

The properties show
that $WCC$ is a metric capable of favoring those communities with a large 
quantity of triangles involving all the vertices of the community. We are 
ensuring, like the informal community definition says, that all the 
vertices forming the community are highly connected among them. This property 
is not fulfilled by other proposed metrics, such as the conductance.
Compared to $WCC$, which is based on triangles, conductance is based 
on the edge cut. 
Minimizing the cut is problematic because the partition consisting of a 
single community containing all the vertices of the graph has the best value, 
and thus it is the optimal community. This makes impossible to design 
algorithms that simply optimize conductance.
Properties~1, 2 and~3 do not apply to conductance, because as soon 
as there is an edge connecting two communities or a single 
vertex connecting to a community, 
joining them into a single community will grant a better conductance value.

\begin{figure}[t]
\centering
\includegraphics[width=1\linewidth]{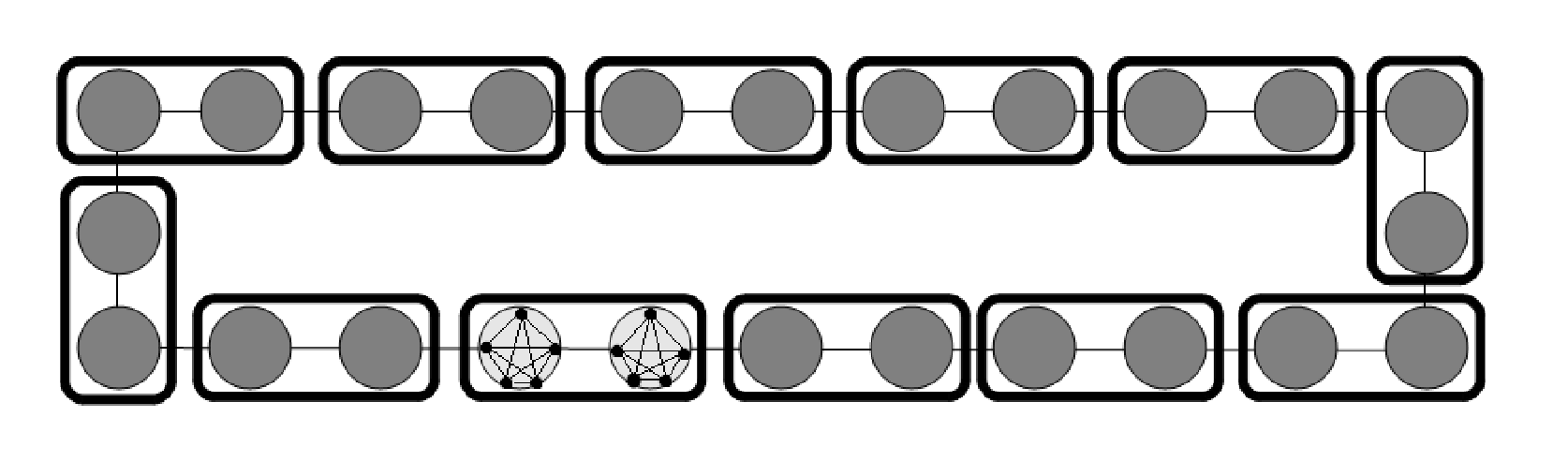}
\caption{Ring with 24 cliques of 5 vertices each (shaded circles). 
Setting each clique as a community has a modularity of 0.8674, but merging
adjacent cliques has modularity 0.8712~\cite{good2010performance}.}
\label{fig:modularity_resolution}
\end{figure}

In the case of modularity, it suffers from resolution
limits~\cite{fortunato2007resolution,good2010performance}.
This resolution problem 
is exemplified in Figure~\ref{fig:modularity_resolution}, where
the optimal communities for modularity are groups of two cliques.
In this example, the communities found by optimal modularity 
contain a bridge, and thus they do not verify 
Property 3. However, the natural communities are the 
groups of five vertices forming cliques, which are the optimal communities
for $WCC$. The $WCC$ value of the five vertices 
clique is one, so having a partition with each 
clique as a community has the maximum $WCC$ value. Furthermore,
is has been shown~\cite{bagrow2012communities} that 
trees, which cannot be considered communities, can have 
arbitrarily large modularity. 
We show that $WCC$ is a metric that sees the communities in a 
local fashion, focusing in the internal density and the connections with
their surroundings instead of the whole graph. Modularity assumes that
graphs are homogeneous, whereas they are not.

\section{Experiments}\label{section:experiments}

In this section, we first select some of the most relevant community 
detection algorithms and analyze them. 
We use synthetically generated graphs from
which we know the communities beforehand, 
to show that $WCC$ favors those community 
detection algorithms that best capture the actual communities. 
We also execute the algorithms on 
graphs from real world data. 
We prove experimentally 
that $WCC$ captures the nature of a community 
by studying the correlation 
of $WCC$ with statistical properties of the communities. 
%

The algorithms used to extract the communities are 
\textit{Infomap} \cite{rosvall2008maps}, which
is based on random walks; \textit{Blondel} \cite{blondel2008fast}, 
which is based on multilevel maximization of modularity
locally; \textit{Clauset} \cite{clauset2004finding}, 
which maximizes the modularity iteratively; 
\textit{Newman} \cite{newman2006modularity}, which maximizes modularity
by exploiting the spectral properties of 
the graph and \textit{Duch} \cite{duch2005community}, which uses
heuristic search based on extremal optimization to optimize the modularity.


We choose Infomap and Blondel because they are the best for detecting
communities in social networks according to~\cite{lancichinetti2009community}.
Clauset, Newman and Duch are chosen because their popularity in the
literature. The implementations of Infomap, Clauset and Blondel are taken from
their authors' web. In the case of Newman and Duch, we have used the Radatools
library~\cite{radatools2011}. Our selection covers a wide range of community
methods to test the validity of $WCC$ 
but does not intend to be an evaluation survey of all community methods. 
Besides, other popular approaches in the literature, such as
~\cite{Lancichinetti2010,ahn2010link,palla2005uncovering} among others, aim 
at overlapping communities which are also out of the scope of this paper.


\subsection{Synthetic Graphs}

In this section, we use synthetically generated 
graphs, where the communities are
known beforehand. 
We build graphs of 10k vertices 
with social network topology with a generator~\cite{lancichinetti2008}. 
We use the default parameters, which
are typical of social networks 
(used also in ~\cite{lancichinetti2009community}). We vary
the mixing factor, which is the percentage of edges 
that connect a vertex with other 
vertices outside the community, from 0.1 to 0.7. 

\begin{figure*}[t]
\begin{minipage}{0.33\linewidth}
\centering
\includegraphics[width=1\linewidth]{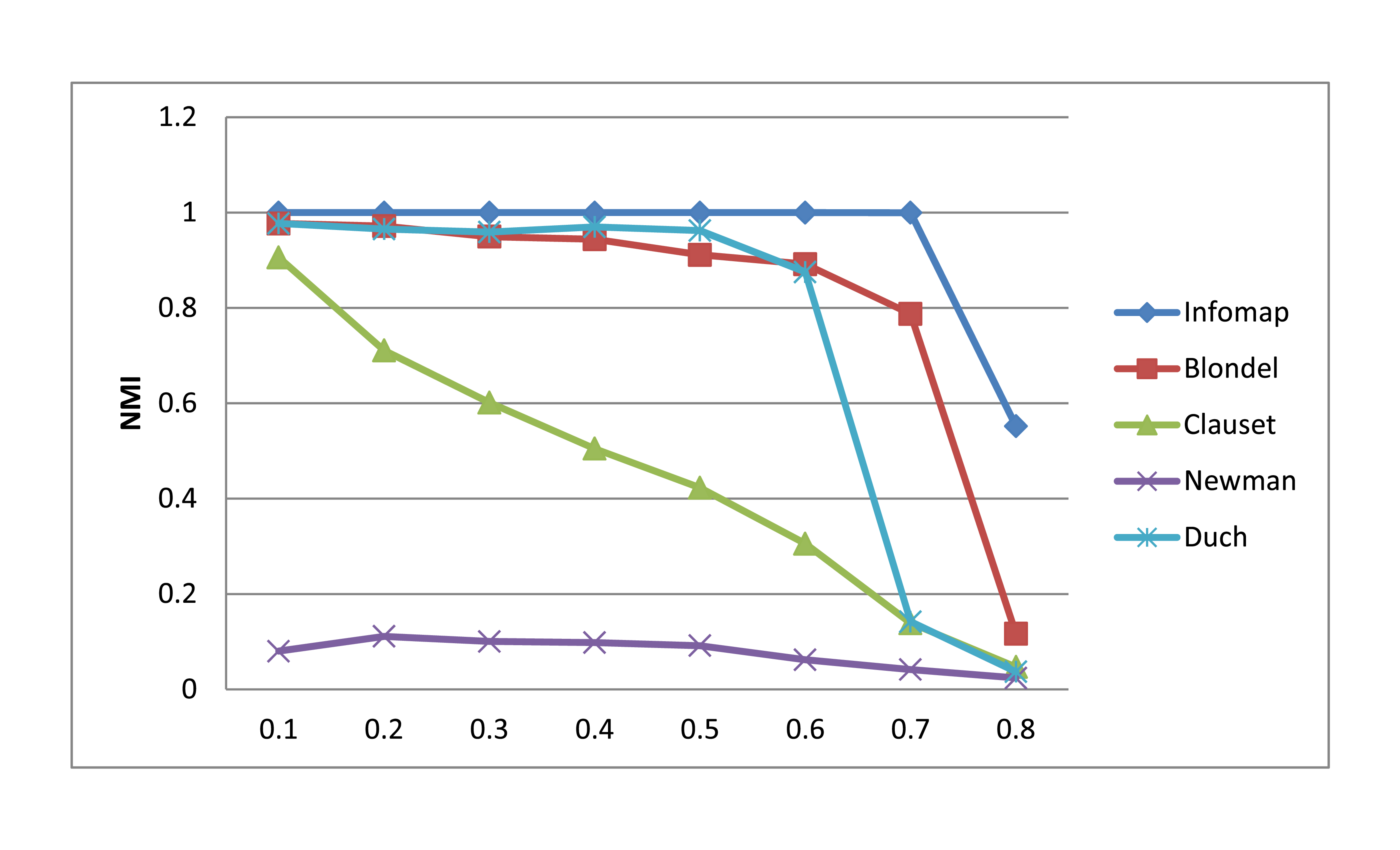}
\end{minipage}
\begin{minipage}{0.33\linewidth}
\centering
\includegraphics[width=1\linewidth]{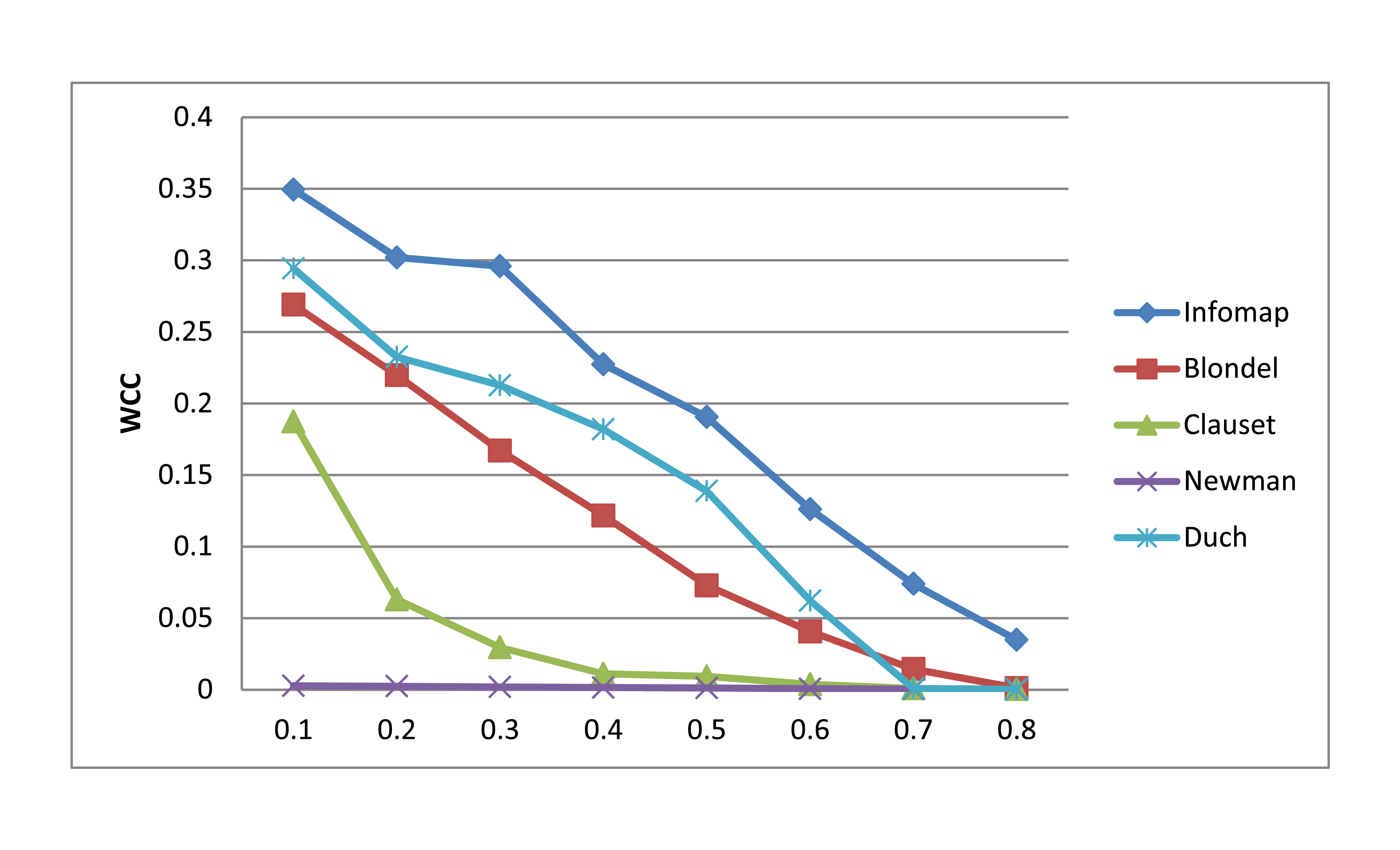}
\end{minipage}
\begin{minipage}{0.33\linewidth}
\centering
\includegraphics[width=1\linewidth]{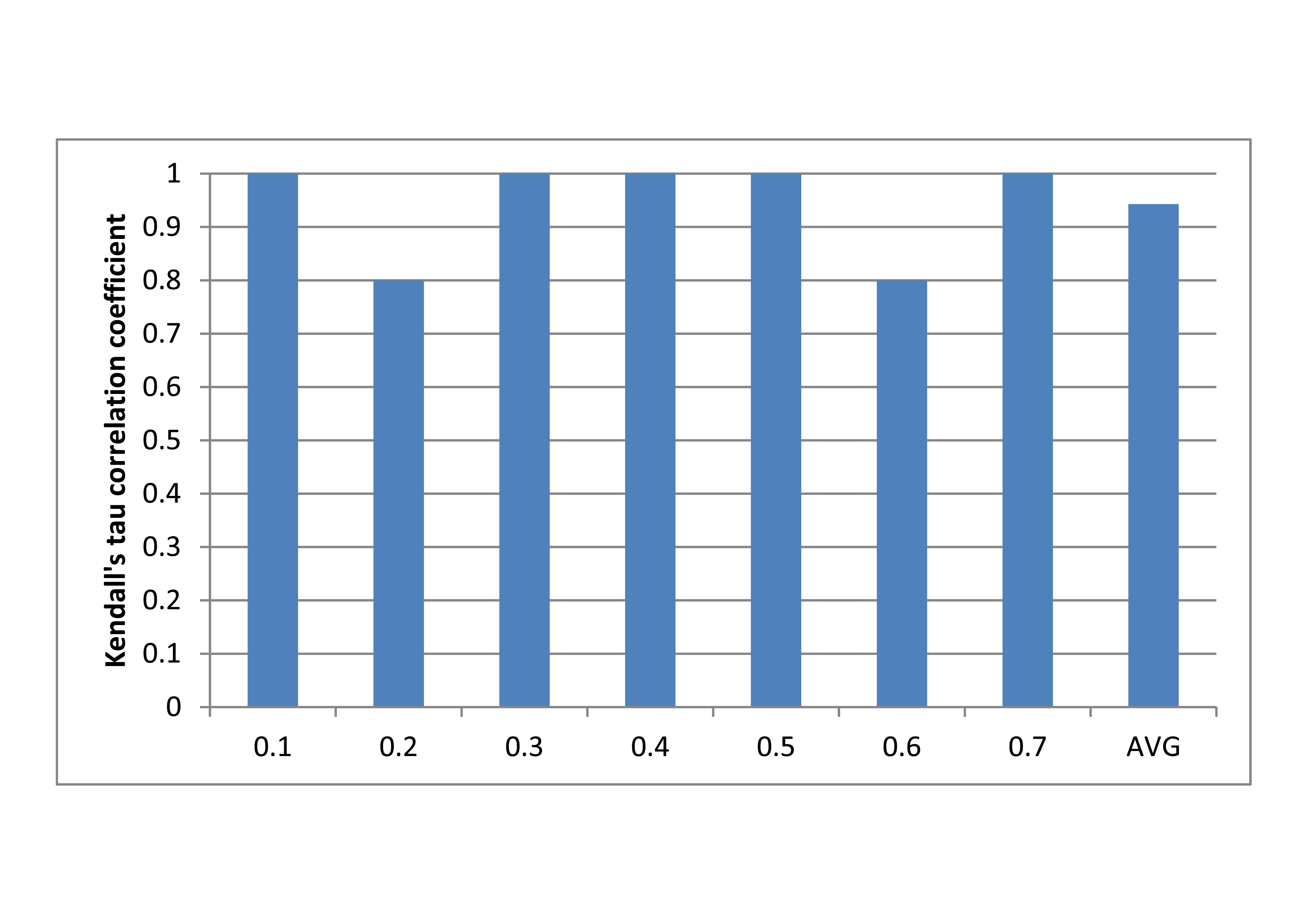}
\end{minipage}
\begin{minipage}{0.33\linewidth}
\centering
(a)
\end{minipage}
\begin{minipage}{0.33\linewidth}
\centering
(b)
\end{minipage}
\begin{minipage}{0.33\linewidth}
\centering
(c)
\end{minipage}
\caption{ (a) NMI value and (b) $WCC$ value for the most relevant state of the art algorithms on synthetic graphs with different
mixing factors. (c) Kendall's tau correlation coefficient between NMI and $WCC$.}
\label{fig:sintetics}
\end{figure*}

The quality of the result for each algorithm is measured by the 
\emph{normalized mutual information} (NMI), which computes the overlap between
the algorithm output and the benchmark~\cite{fred2003robust}, 
and it is shown in Figure~\ref{fig:sintetics}(a).
We see that Infomap stands as the best algorithm, 
followed by Blondel and Duch at a small distance. We
see that, the larger the mixing factor, the higher 
the difficulty to correctly find
the communities by the algorithms.

On the other hand, in Figure~\ref{fig:sintetics}(b) we show the $WCC$ for
each algorithm. In this case, the best algorithm is again Infomap.
Blondel and Duch perform slightly worse than Infomap, and Clauset and
Newman stand as the worst algorithms of all in terms of $WCC$.
We observe that $WCC$ is a good model for the communities built by the benchmark.
Those algorithms with high NMI also have high $WCC$.

We quantify the correlation of both metrics in Figure~\ref{fig:sintetics}(c).
We apply the Kendall's rank correlation
coefficient~\cite{borowski2005collins}, which compares two relations
of order. A value
of 1 indicates that the two metrics are fully correlated, 
while 0 indicates that no correlation is found. 
If the two metrics are inversely correlated the value is -1.

In Figure~\ref{fig:sintetics}(c), we observe that the agreement of both
metrics is excellent.
For all mixing factors, except
for 0.2 and 0.6, the Kendall's tau correlation coefficient is 1, which means
that the correlation is perfect. Only one pairwise comparison for 0.2 and
0.6 mixing factors is reversed, which correspond to the pair Blondel-Duch
(see Figures ~\ref{fig:sintetics} (a) and (b)). However, in both cases the
difference in NMI between both methods is tiny (less than 1\%), and thus, it
is difficult to discern which community partition is better. Moreover, for all
mixing factors the kendall significance test (significance 0.05) concludes
that there is statistical evidence that both variables are correlated.

Overall, the average Kendall correlation is 0.94, which is very high.
In other words, the synthetic community generation procedure,
which is not based on triangles, generates communities that match our community
definition based on $WCC$. Therefore, we conclude that $WCC$
captures community structure effectively and that $WCC$ is an adequate 
indicator of the quality of the communities found in a graph. 

\begin{table}[t]
\centering
{\footnotesize
\begin{tabular}{|m{1.5cm}|m{1.2cm}|m{1.2cm}|m{1.2cm}|m{1.2cm}|}
\hline
& {ArxivCit} & {ArxivCol} & {Enron} & {Slashdot} \\
\hline
Vertices & 27,769  & 18,771 & 36,692 &  82,168\\
\hline
Edges & 352,285 & 198,050 & 183,831 &  504,230\\
\hline 
Avg.degree & 25.37 & 21.1 & 10.02 &  12.27\\ 
\hline
Max.degree & 2,468  & 504 & 1,383 &  2,552\\ 
\hline
\end{tabular}}
\caption{The real world graphs used for testing.}
\label{tab:real_graphs}
\end{table}

\subsection{Real World Networks}

In this section, we show that there is a correllation 
between communities with good
$WCC$ values and good statistics. 
We study the following measures:
\textit{triangle density}, which is the number of
internal triangles in the community divided 
by the total number of possible internal triangles; 
the \textit{average inverse edge cut}, which is, the average number
of neighbors of a vertex that belong to the same community 
divided by the total number 
of neighbors; the \textit{average edge density}, 
which is the average number of neighbors that a vertex has in the community 
divided by the total number of members of the community;
the \textit{ modularity}; the \textit{conductance}; the \textit{normalized diameter}, 
which is the 
diameter of the community divided by the logarithm of its size;
the \textit{ bridge ratio}, which is the percentage of edges 
in a community that are bridges; and the vertex \textit{size} of the communities.

We create a pool of communities by running 
the community detection algorithms on four real world networks,
covering different aspects of real world
data\footnote{Downloaded from SNAP 
(\url{http://snap.stanford.edu}). We cleaned the
original graphs by removing the self loops.
}. 
ArxivCit is a citation network, ArxivCol represents the collaborations
between scientists, Enron is derived from email communications and Slashdot
is extracted from a website social network.
Table~\ref{tab:real_graphs} summarizes the graph properties.

\begin{figure*}[t!]
\begin{minipage}{0.33\linewidth}
\centering
\includegraphics[scale=0.2,angle=-90]{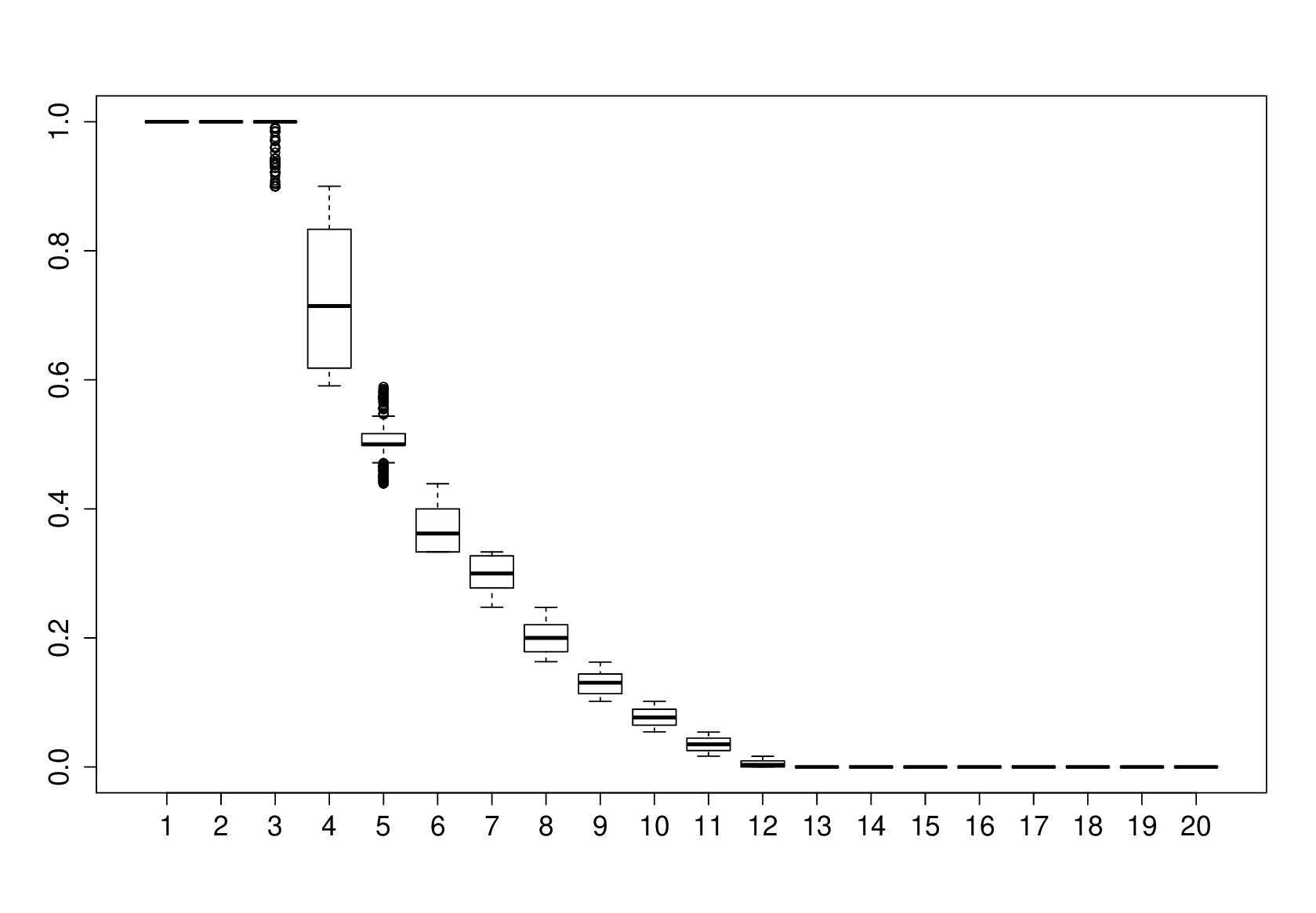}
\end{minipage}
\begin{minipage}{0.33\linewidth}
\centering
\includegraphics[scale=0.2,angle=-90]{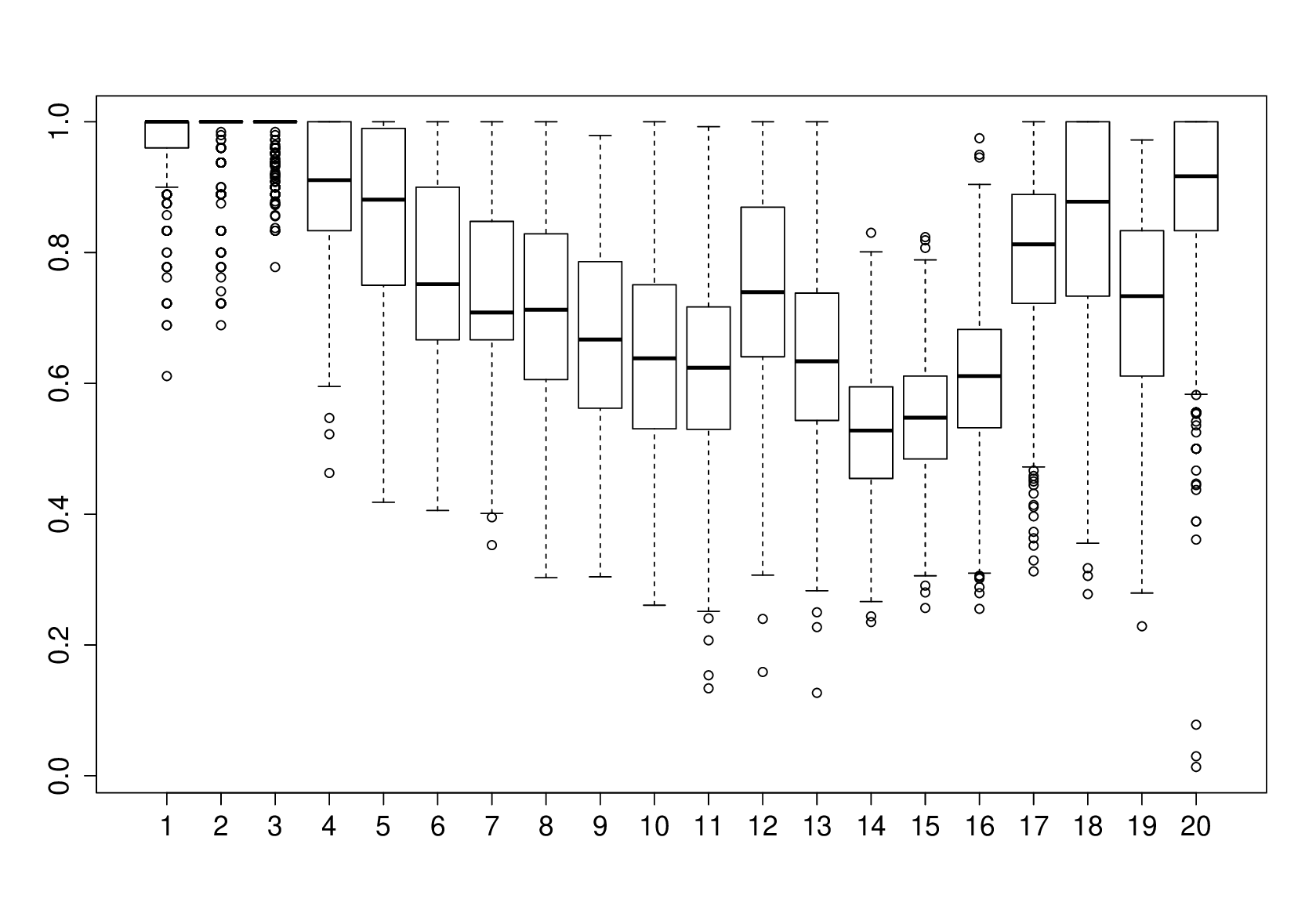}
\end{minipage}
\begin{minipage}{0.33\linewidth}
\centering
\includegraphics[scale=0.2,angle=-90]{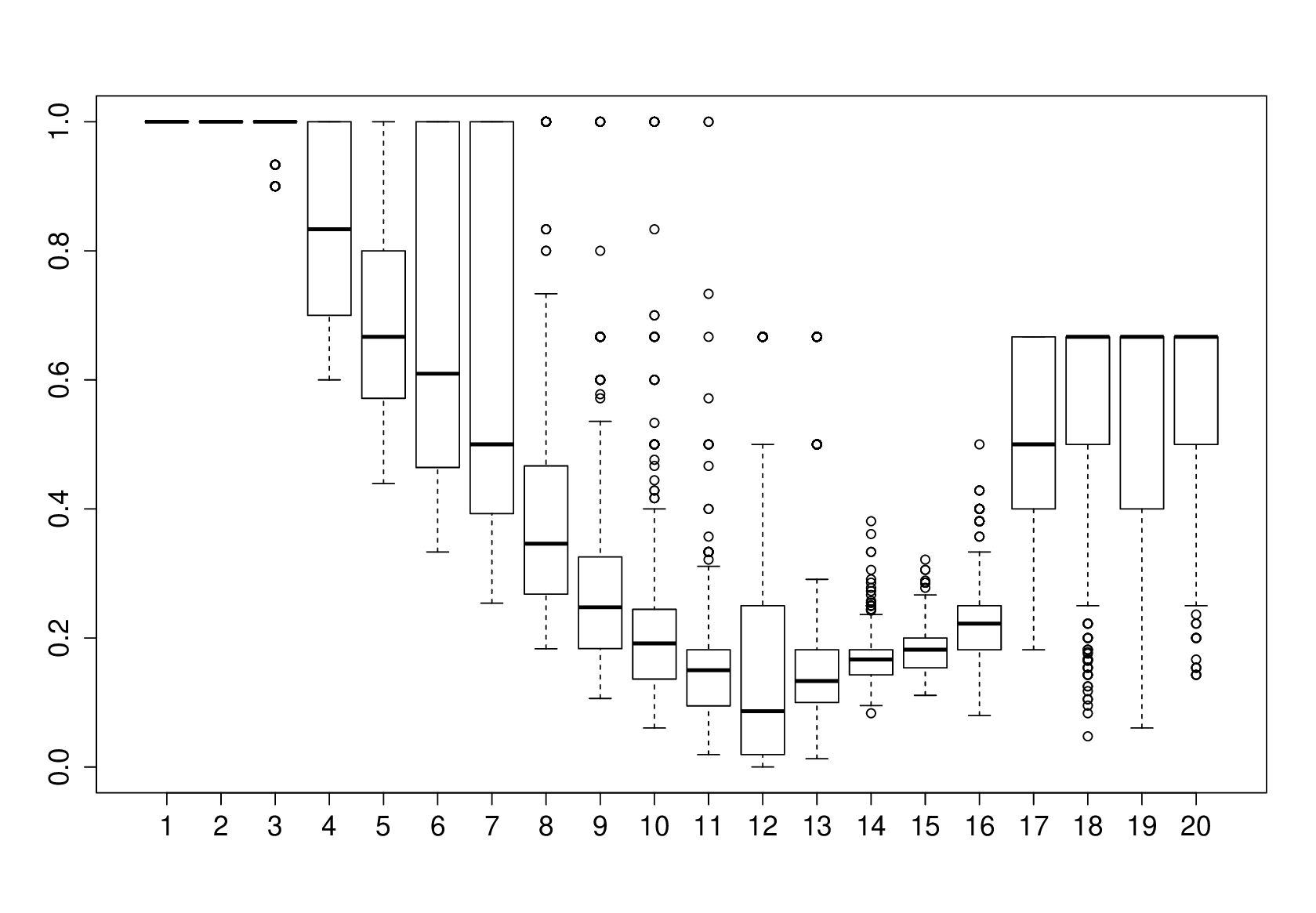}
\end{minipage}

\begin{minipage}{0.33\linewidth}
\centering
(a) WCC
\end{minipage}
\begin{minipage}{0.33\linewidth}
\centering
(b) Average inverse edge cut
\end{minipage}
\begin{minipage}{0.33\linewidth}
\centering
(c) Average edge density
\end{minipage}

\begin{minipage}{0.33\linewidth}
\centering
\includegraphics[scale=0.2,angle=-90]{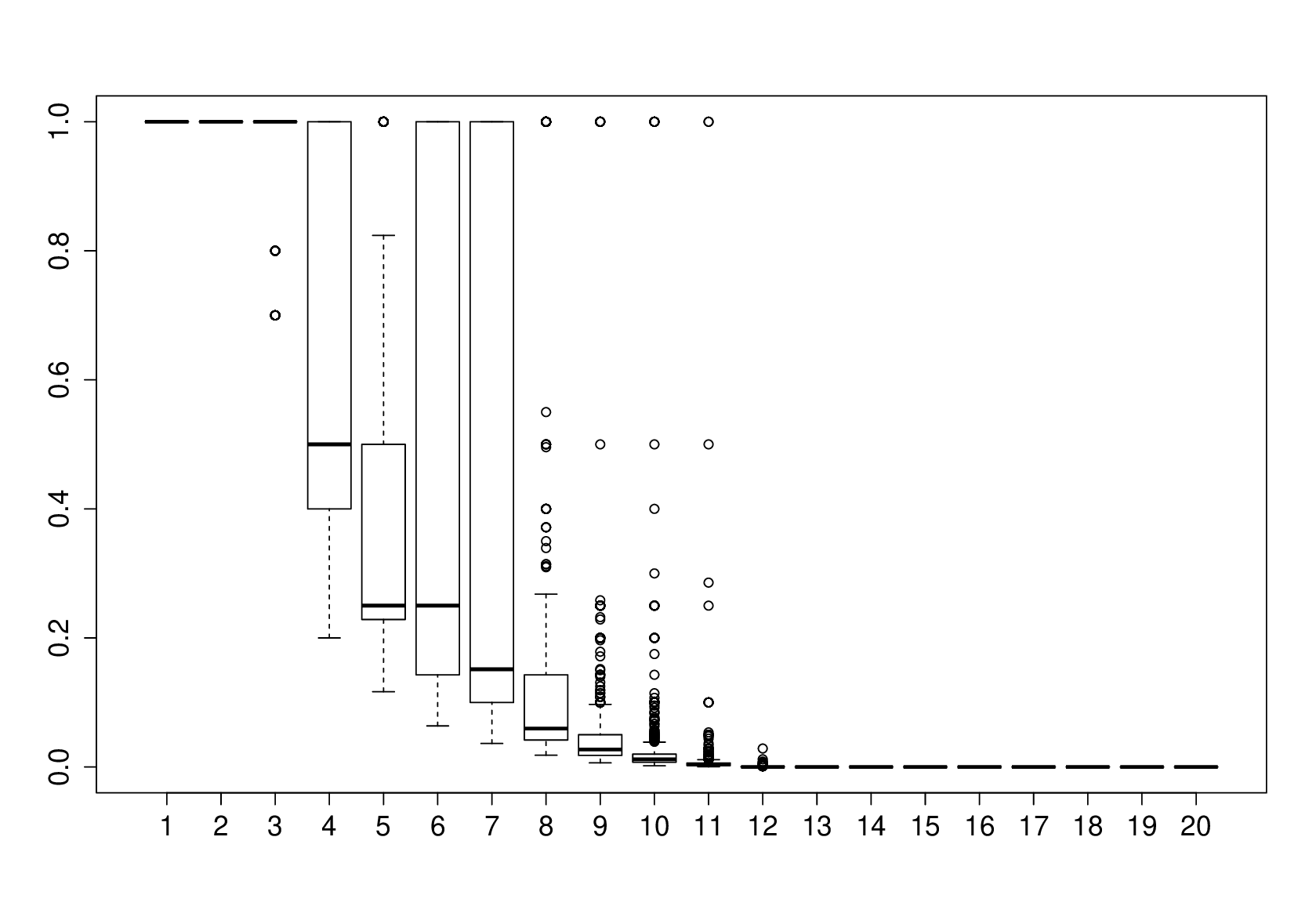}
\end{minipage}
\begin{minipage}{0.33\linewidth}
\centering
\includegraphics[scale=0.2,angle=-90]{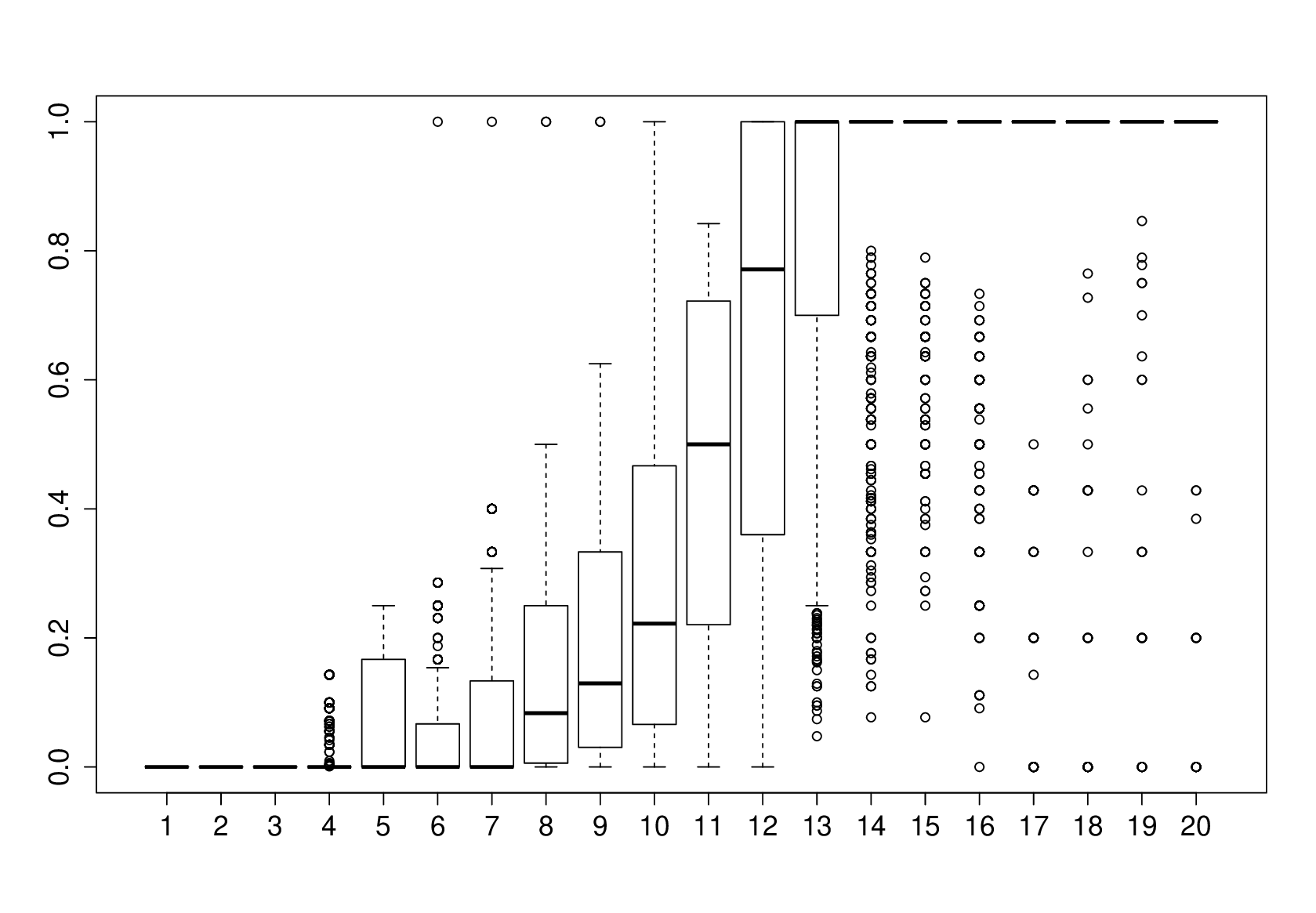}
\end{minipage}
\begin{minipage}{0.33\linewidth}
\includegraphics[scale=0.2,angle=-90]{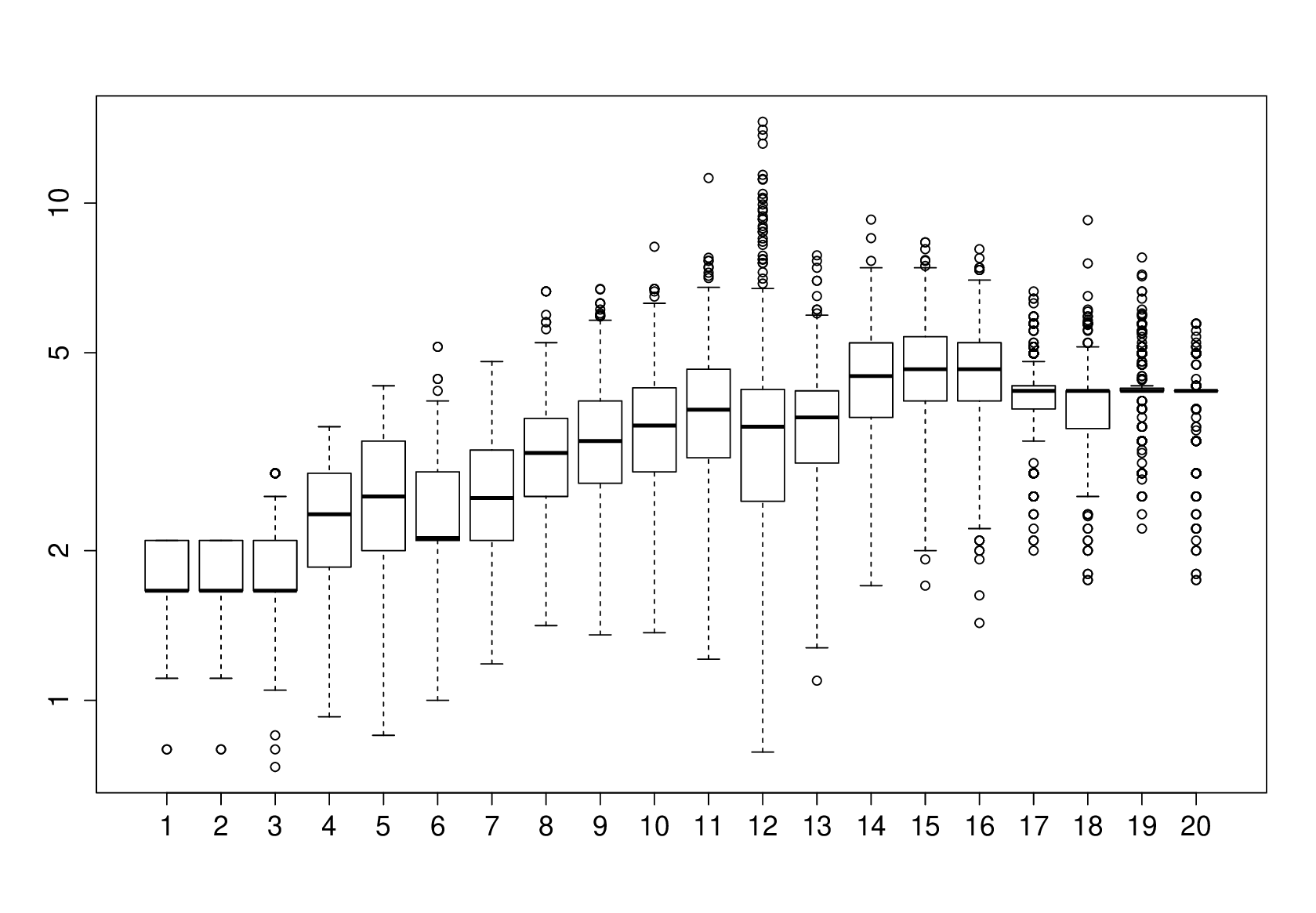}
\centering
\end{minipage}

\begin{minipage}{0.33\linewidth}
\centering
(d) Triangle density
\end{minipage}
\begin{minipage}{0.33\linewidth}
(e) \centering
Bridge ratio
\end{minipage}
\begin{minipage}{0.33\linewidth}
\centering
(f) Diameter
\end{minipage}

\begin{minipage}{0.33\linewidth}
\centering
\includegraphics[scale=0.2,angle=-90]{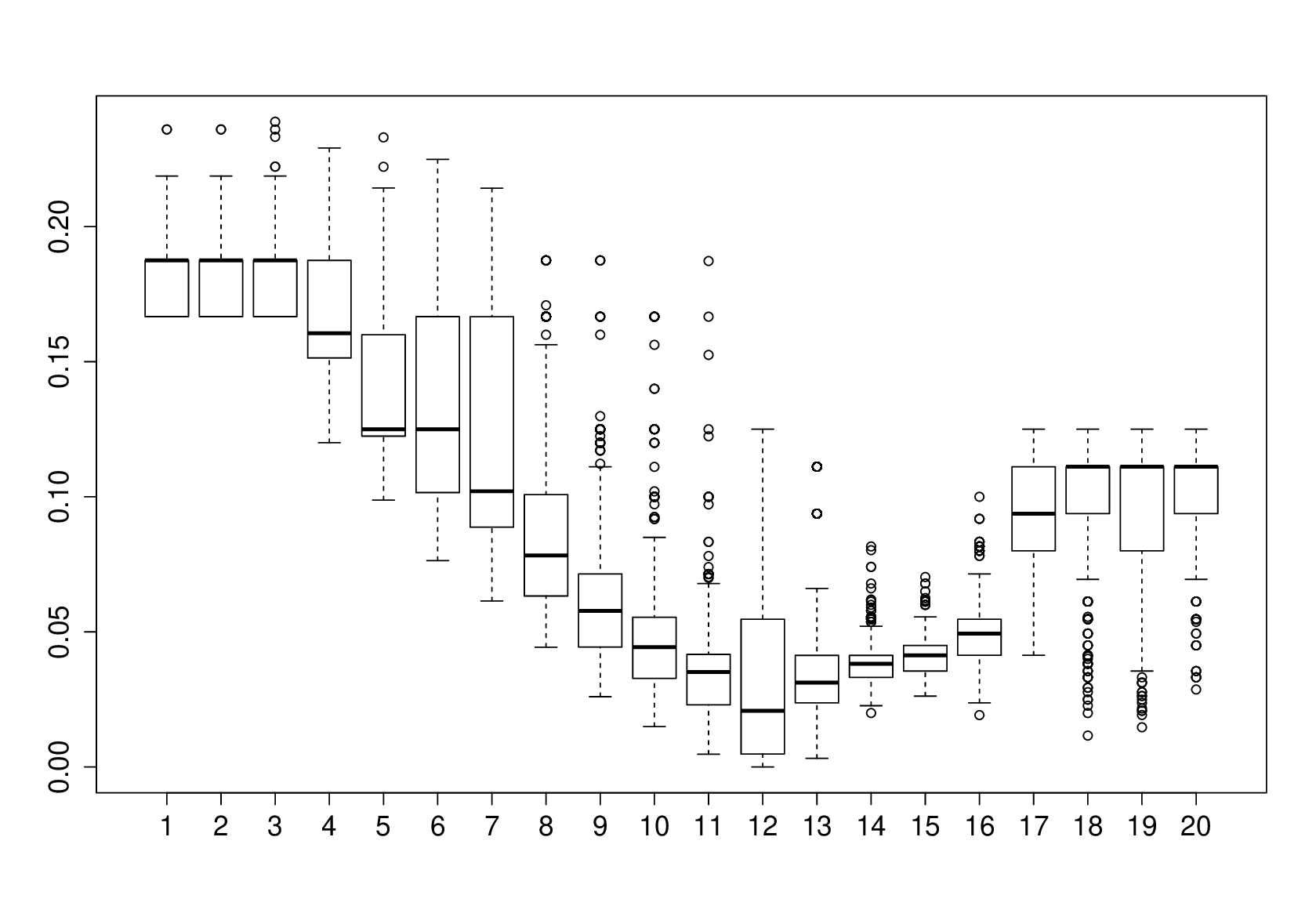}
\end{minipage}
\begin{minipage}{0.33\linewidth}
\centering
\includegraphics[scale=0.2,angle=-90]{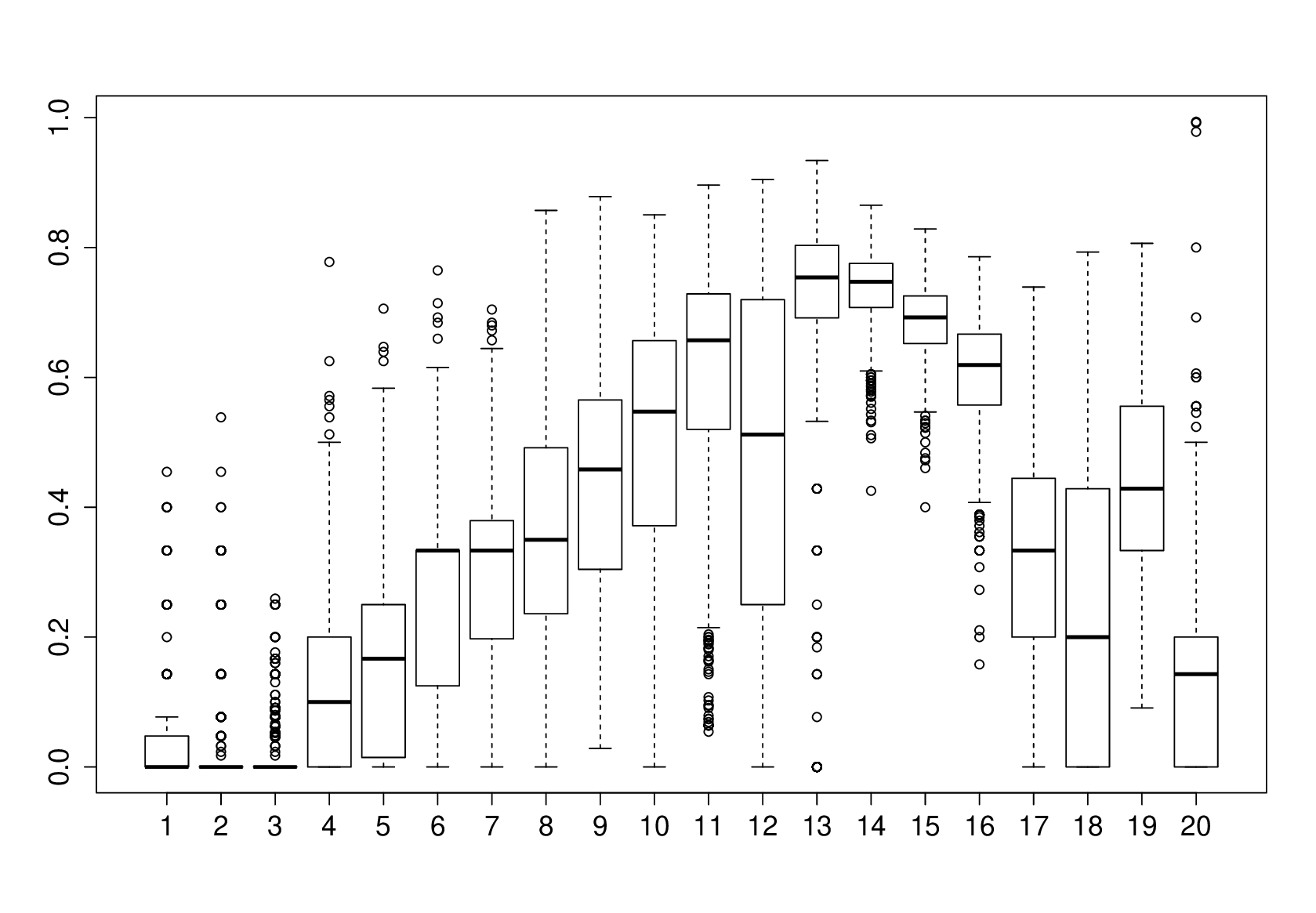}
\end{minipage}
\begin{minipage}{0.33\linewidth}
\centering
\includegraphics[scale=0.2,angle=-90]{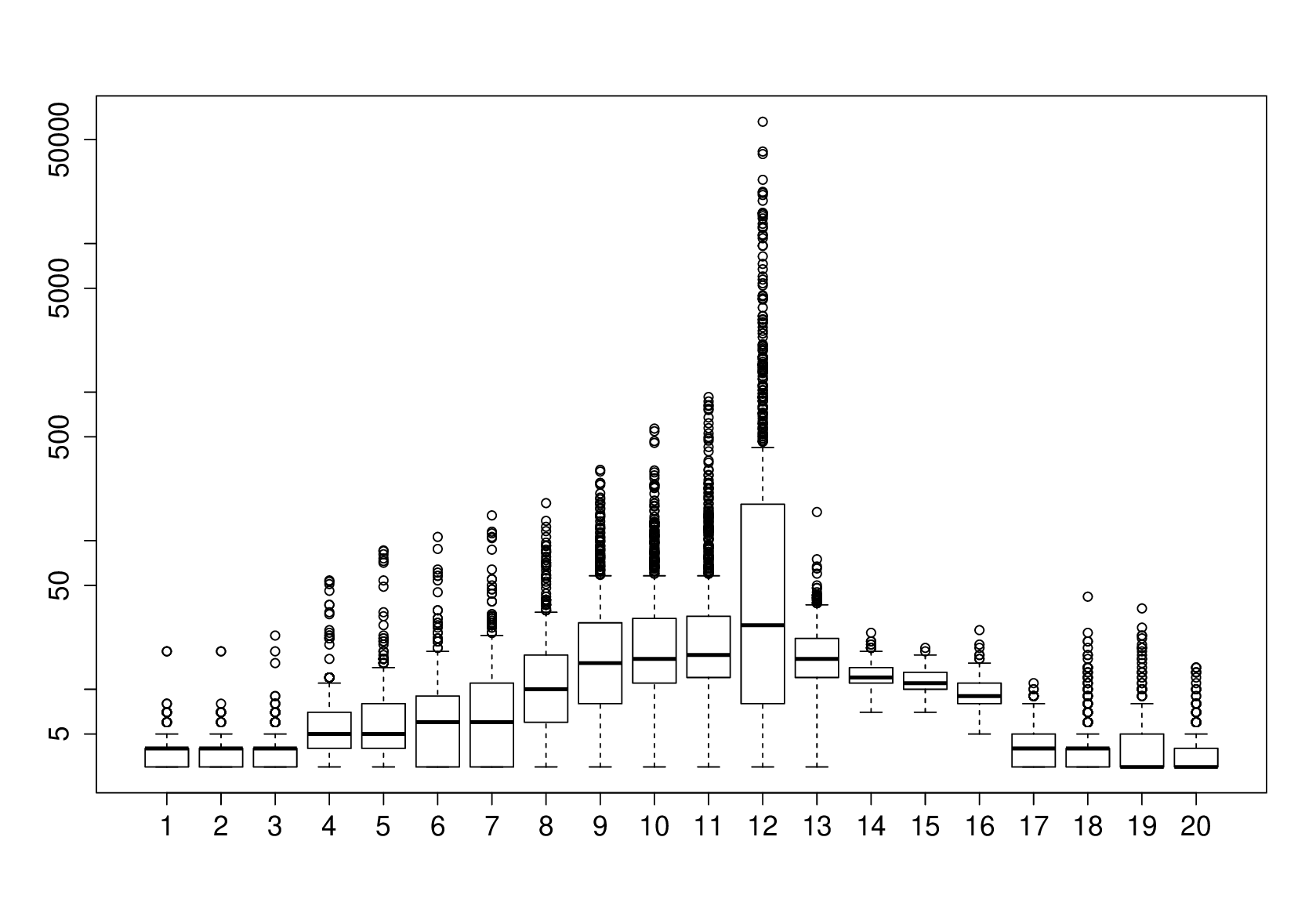}
\end{minipage}

\begin{minipage}{0.33\linewidth}
\centering
(g) Modularity
\end{minipage}
\begin{minipage}{0.33\linewidth}
\centering
(h) Conductance
\end{minipage}
\begin{minipage}{0.33\linewidth}
\centering
(i) Size
\end{minipage}

\caption{Statistics of communities from real world networks in 20 groups sorted by WCC.}
\label{fig:statistics_com}
\end{figure*}

We sorted all the communities obtained by the five algorithms 
by their $WCC$ value decreasingly. Then, we divided the communities into 20 groups
in steps of five percentiles according to their $WCC$ and plotted for 
these 20 groups their correponding statistics in Figure~\ref{fig:statistics_com}. 
In all the charts, the x axis represents the group identifier (e.g. the leftmost 
group is always the 95 percentile that contains the top 5\% communities in terms of their
$WCC$) while the y axis shows the corresponding statistical value. 
The communities of size one and two, 
are ommited since their $WCC(S)$ value is always zero. As shown in Figure~\ref{fig:statistics_com}(a), the
leftmost communities have high $WCC$ values, and the rightmost communities have the lowest $WCC$ values.

Broadly speaking, we observe two sections: from groups 1 to 12, the trends
for all statistics show that communities with higher $WCC$ have 
better properties; from groups 13 to 20 this trend apparently
changes in some statistics. We focus first on groups 1-12 and we analyze
groups 13-20 below. 

\textit{Groups 1-12:}
In Figure~\ref{fig:statistics_com}(b), we see that the larger the $WCC$, 
the smaller the edge cut so, the number of external connections of the
community decreases. On the other hand, in Figure~\ref{fig:statistics_com}(c) we see that the larger the $WCC$
of a community, the larger the internal density of edges. While these two
characteristics are a good starting point to identify 
a good community, they do not
imply an internal structure which is shown in Figure~\ref{fig:statistics_com}(d):
the larger the $WCC$ of a community, the larger its triangle density.
These transitive relations between the vertices (Property 1) indicate a good social
structure of the communities.

Figure~\ref{fig:statistics_com}(e) shows how bridges penalize $WCC$. 
A large percentage of bridges is a symphtom of the presence of 
whiskers or treelike structures, which are inherently
sparse and hence do not have the type of internal structure that one 
would expect from a community. We note that communities that contain bridges 
are not the optimal communities because of Property 3.
A small diameter is another feature that any good community should have. In
Figure~\ref{fig:statistics_com}(f) we see that
large $WCC$ implies smaller diameters for the communities. This means 
that any vertex in the community is close to any
other vertex, which translates to denser communities. 

In Figure~\ref{fig:statistics_com}(g-h) we compare $WCC$ 
with the two other metrics in the state of
the art: modularity and conductance. We see that there is 
a correlation between communities with good $WCC$ values
and modularity and conductance (for conductance, the lower, the better). 
Finally, in Figure~\ref{fig:statistics_com}(j), 
we show the sizes of the communities. 

\textit{Groups 13-20:} We see that there is a change on the 
trend for some statistics for those groups that have $WCC$ close to 0.
This behavior can be explained by
Figures~\ref{fig:statistics_com}(d) and (e). These figures
reveal that the communities after group 13
are treelike: almost all the edges in the community are
bridges, and communities hardly contain triangles. Therefore, 
such structures cannot be accepted as good communities.
Although some communities in group 13-20 are isolated (and thus 
have good conductance), we note that this is not a sufficient condition to be 
good communities. For example, most communities in 
groups 17-20 are trees with three vertices, which have a good conductance.
As described in~\cite{bagrow2012communities}, 
tree like networks can have high modularity and hence, algorithms maximizing
it can lead to misleading results (Figure~\ref{fig:statistics_com}(g)).

\begin{figure}
\centering
\includegraphics[width=0.90\linewidth]{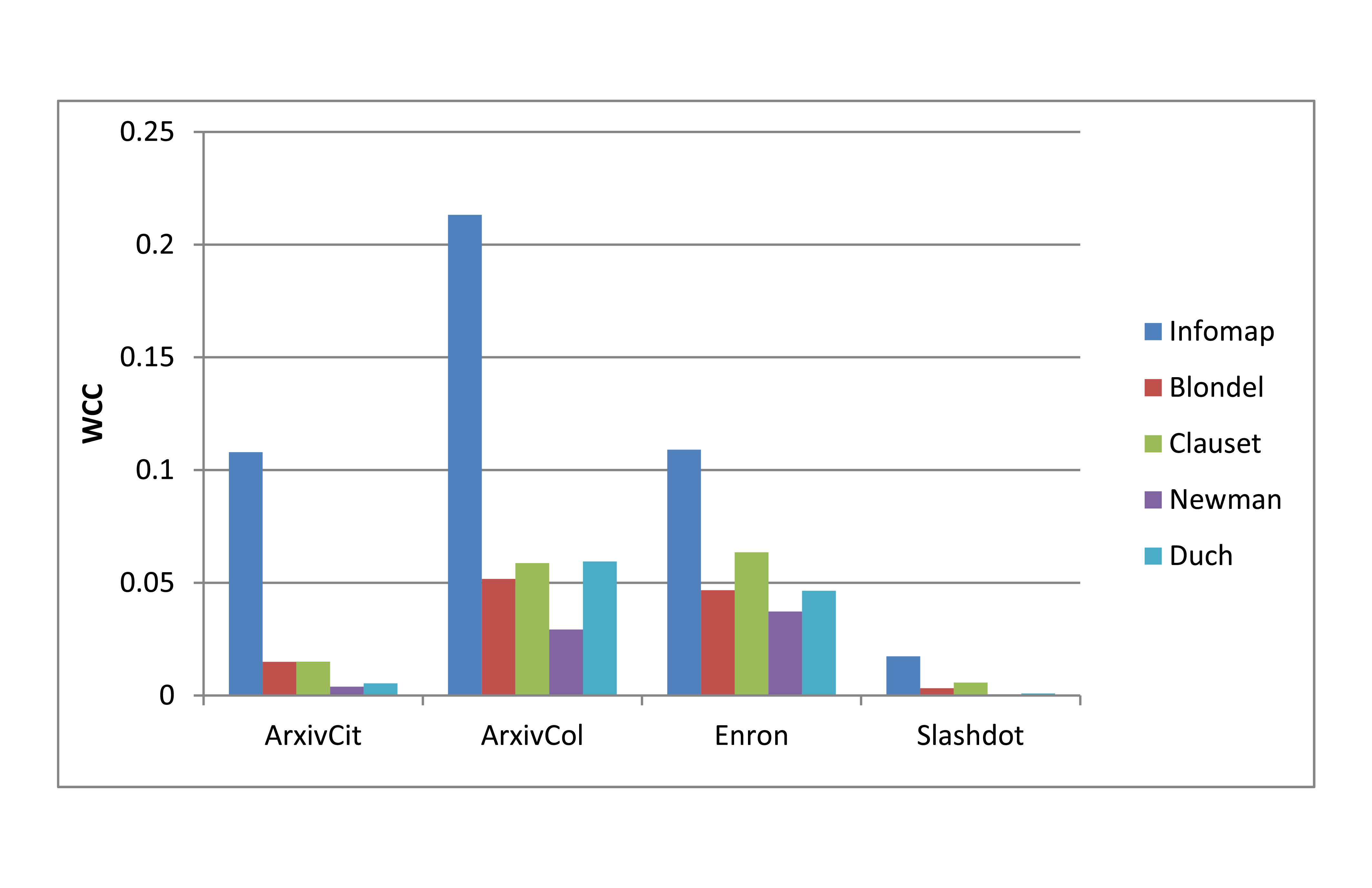}
\caption{$WCC$ for communities from real graphs.}
\label{fig:executions}
\end{figure}

Finally, in Figure~\ref{fig:executions} we compare the different algorithms used
in terms of $WCC$. We see that the results obtained are similar to those
obtained with synthetic graphs, with Infomap outperforming the rest of the
algorithms. However, we see that in this case, Clauset performs at a level
comparable to Blondel and Duch or even better. This might indicate that synthetic
graphs fail at accurately represent the inhomogeneities and noise present in
real graphs, so using this graphs only when evaluating the
quality of community detection algorithms can derive to misleading results.

\section{Conclusions and Future Work}\label{section:conclusions}

Although different metrics have been previously proposed to evaluate 
the quality of a graph partition into communities for social networks, these  
fulfill only partially the concept of community. Even the most popular 
metrics applied in the state of the art (modularity and conductance) fail 
at meeting some minimal properties expectable 
for a social network communities,
such as dealing with bridges, vertex cut, scalability on
the community formation or imposing a minimal 
internal structure such as the triangle.
The reason for this is that the current metrics are based on the informal
definition of a community, which is unable to fully
capture the community concept for social networks 
by its own. We conclude that the concept of community is strongly 
dependent on the domain of the graphs being analyzed, which is something that
the current metrics do not take into account. This suggests that the definition
of a minimal set of properties extending 
this formal definition is required, in order to exploit the 
inherent characteristics of the type of graph being analyzed and its semantics 
(social networks in our paper).

In this paper, we proposed $WCC$, which compares
the quality of two graph community partitions. 
Such a metric captures the community concept by meeting the enumerated
minimal properties, enabling
to distinguish a good from a bad community partition automatically. 
Then, it is possible to compare the quality of algorithms or even design
efficient community computing algorithms based on $WCC$. We have shown 
experimentaly that communities with good $WCC$ values
are dense, have small edge cuts, have transitive
relations without bridges and small diameters. We have also shown that looking
only at the internal density and small edge cuts does not guarantee well defined
communities with internal structure, since it can lead to treelike communities.


Regarding the future work, an interesting problem 
related to that discussed in the paper is the 
location of overlapped communities in the graph.
Some graph patterns, such as cut vertices, can be naturally modeled as the
overlap of two communities. The overlapped community problem, similarly to
the non-overlapped case, has similar deficiencies 
in the sense that there is no formalization of a minimal
set of properties that a metric should fulfill.
Therefore, our work will continue toward extending the 
community definition concepts and $WCC$ to
the detection of overlapping communities.

\section{Acknowledgments}
The members of DAMA-UPC thank the Ministry of Science and Innovation of Spain and Generalitat de Catalunya, 
for grant numbers TIN2009-14560-C03-03 and GRC-1187 respectively, and IBM CAS Canada Research for their 
strategic research grant. Josep M.Brunat thanks he Ministry of 
Science and Innovation of Spain and Generalitat de Catalunya for grant numbers MTM2011-24097 and DGR-2009SGR1040
respectively. David Dominguez-Sal thanks the Ministry of Science and Innovation of Spain for the grant Torres Quevedo PTQ-11-04970.

\bibliographystyle{abbrv}
 {\small
\bibliography{cikm12}
 }

\begin{appendix}

\setcounter{teorema}{0}
\setcounter{corolari}{0}
\setcounter{proposition}{0}
\setcounter{equation}{0}

\section{Proof of Proposition 1}\label{appendix:formal_proof_init}


\begin{proof}
(i)  This is a consequence of the inequalities 
$t(x,S)\le t(x,V)$ and 
\begin{footnotesize}
\begin{eqnarray}
vt(x,V)&=& vt(x,S)+vt(x,V\setminus S)\\ \label{dsf1}
      &\le& |S\setminus\{x\}|+vt(x,V\setminus S).\label{dsf2}
\end{eqnarray} 
\end{footnotesize}
(ii) If $WCC(x,S)=0$, then at least one of the following three identities
holds: $t(x,V)=0$, 
$vt(x,V)=0$, and $t(x,S)=0$. Now, each one of these conditions implies
$t(x,S)=0$. Reciprocally, by definition, if $t(x,S)=0$, then $WCC(x,S)=0$. 

(iii) Assume $WCC(x,S)=1$. By (ii), $t(x,S)\ne 0$. Hence, there exists
an edge $\{y,z\}: y \in (S\setminus\{x\})$ and $z \in (S\setminus\{x\})$ forming triangle with $x$. Then
$|S\setminus\{x\}|\ge 2$. As the two fractions defining $WCC(x,S)$ are $\le
1$, the condition $WCC(x,S)=1$ implies that both fractions are $1$.  
The
condition $t(x,V)=t(x,S)$ is equivalent to $vt(x,V\setminus S)=0$.
Since $WCC(x,S)=1$, the inequality~(\ref{dsf2}) is an
equality, and we have $vt(x,S)=|S\setminus\{x\}|$. 

Reciprocally, the condition $vt(x,V\setminus S)=0$ implies
$t(x,S)=t(x,V)$ and $vt(x,S)=vt(x,V)$.  As
$vt(x,V)=vt(x,S)=|S\setminus\{x\}\ge 2$, we have that both fractions
in the definition of $WCC(x,S)$ have denominator $\ne 0$ and both
fractions are $1$.  Therefore, $WCC(x,S)=1$.
\end{proof}

\section{Proof of Proposition 2}\label{appendix:formal_proof_B}


\begin{proof}
The proofs are a consequence of Proposition~\ref{proposition:proposicionsNode}.
(i) Since $0 \leq WCC(x,S) \leq 1$  $ \forall x\in S$, then $0 \leq WCC(S) \leq 1$. 
(ii) $WCC(S) = 0$ implies that $\forall x \in S$ $WCC(x,S)=0$. Since the condition for $WCC(x,S) = 0$ is that $t(x,S) = 0$,
       then $WCC(S)=0$ implies that $S$ has no triangles.
(iii) $WCC(S) = 1$ implies that for all $x \in S$ $WCC(x,S)=1$. This implies that does not exist a vertex $x \in S$ that $t(x,V\setminus S) \neq 0$ and $vt(x,S)=|S\setminus \{x\}|$.
Thus, all the vertices $x\in S$ form triangles only with and with all the other vertices in $S$, which implies having an edge with all the vertices in $S$, and hence
forming a clique.
\end{proof}

\section{Proof of Theorem 1}\label{appendix:primer_teorema}


\begin{proof}
Let $N$ be the set of neighbors of $v$.

(i) For $x\in V$, we have $WCC(x,V)=vt(x,V)/r$. Now,
\begin{footnotesize}	
$$
vt(x,V)=\left\{\begin{array}{ll}
(r-1)p & \mbox{if } x\in V_r\setminus N;\\
(r-1)p+1 & \mbox{if } x\in N;\\
d        & \mbox{if } x\in \{v\}.
\end{array}\right.
$$ 
\end{footnotesize}

Then 
\begin{footnotesize}
\begin{align*}
(r+1)WCC(\mathcal{P}_1)=&(r-d)\frac{(r-1)p}{r}+d\frac{(r-1)p+1}{r}+\frac{d}{r}\\
=&(r-1)p+2\frac{d}{r}.
\end{align*}
\end{footnotesize}

(ii)
For $x\in V_r\setminus N$,
\begin{footnotesize}
$$
WCC(x,V_r)=\frac{vt(x,V)}{r-1}
        =\frac{(r-1)p}{r-1}=p.
$$
\end{footnotesize}
For $x\in N$, we have
\begin{footnotesize}
\begin{eqnarray*}
t(x,V_r)&=&{r-1\choose 2}p^3;\\
t(x,V)&=&{r-1\choose 2}p^3+(d-1)p;\\
vt(x,V)&=&(r-1)p+1;\\
|V_r\setminus\{x\}|+vt(x,V\setminus V_r)&=&(r-1)+1=r.
\end{eqnarray*}
\end{footnotesize}
Moreover, $WCC(v,\{v\})=0$. Then,
\begin{footnotesize}	
$$
(r+1)WCC(\mathcal{P}_2)=(r-d)p+\frac{d}{r}
        \cdot\frac{((r-1)p+1)(r-1)(r-2)p^2}{(r-1)(r-2)p^2+2(d-1)}.
$$
\end{footnotesize}

(iii) We have,
\begin{footnotesize}
\begin{align*}
(r+1) & \left(WCC(\mathcal{P}_1)-WCC(\mathcal{P}_2)\right)
= p(d-1)+2\frac{d}{r}\\
&-\frac{d}{r}\frac{((r-1)p+1)(r-1)(r-2)p^2}{(r-1)(r-2)p^2+2(d-1)},
\end{align*}
\end{footnotesize}
and the condition $WCC(\mathcal{P}_1)-WCC(\mathcal{P}_2)>0$ 
is equivalent to the condition
\begin{footnotesize}
\begin{equation}
\label{segongrau}
ad^2+bd+c>0,
\end{equation}
\end{footnotesize}
 where
\begin{footnotesize} 
\begin{align*}
a= & 2(2+pr), \\
b= & p^2(p+1)r^2-p(3p^2+3p+4)r+2p^3+2p^2-4, \\
c= & -p^3r^3+3p^3r^2+2p(1-p^2)r.
\end{align*}
\end{footnotesize}
For short, let we denote
by $O(r^n)$ a polinomial expression of degree at most $n$.
Then, the greatest solution of~(\ref{segongrau}) is, 
$$
d_2=\frac{-p^2(1+p)r^2+O(r)+\sqrt{p^4(p^2+2p+9)r^4+O(r^3)}}{4(2+pr)}
$$  
and we get
\begin{footnotesize}
\begin{eqnarray*}
\lim_{r\to +\infty}\frac{d_2}{r}
&=&\frac{-p^2(1+p)+p^2\sqrt{p^2+2p+9}}{4p}\\
&=&p\frac{\sqrt{p^2+2p+9}-(1+p)}{4}.
\end{eqnarray*}
\end{footnotesize}
Thus, for a large enough $r$, the condition 
\begin{footnotesize}
$$
d>rp\left(\sqrt{p^2+2p+9}-(1+p)\right)/4,
$$
is equivalent to
$WCC(\mathcal{P}_1)>WCC(\mathcal{P}_2)$. 
\end{footnotesize}
\end{proof}

Note that the function $p\mapsto p\left(\sqrt{p^2+2p+9}-(1+p)\right)/4$ is
increasing in $p$.  A greater value of $p$ means a greater cohesion in
$G$, and then a greater value of $d/r$ is needed for $WCC(\mathcal{P}_1)$ being
greater than $WCC(\mathcal{P}_2)$.

In the case of Corollary~\ref{corollary:clique}, $p=1$, thus 
$d>\sqrt{3}-1/2=0.37$.

\section{Proof of Theorem 2}

\textsc{Proof.} Let $S=S_1\cup S_2$.
For $x\in S_i, i\in\{1,2\}$ we have $t(x,S_i)=t(x,S)$, $vt(x,V\setminus S_i)=vt(x,V\setminus S)$
and $|S_i\setminus \{x\}|<|s\{x\}|$. Then,
\begin{footnotesize}    
\begin{eqnarray*}
WCC(x,S) & = & \frac{t(x,S)}{t(x,V)} \cdot \frac{vt(x,V)}{|S \setminus \{x\}|+vt(x,V \setminus S)}\\ 
	 & < & \frac{t(x,S_i)}{t(x,V)} \cdot \frac{vt(x,V)}{|S_i \setminus \{x\}|+vt(x,V \setminus S_i)}\\
	 & = & WCC(x,S_i).		  
\end{eqnarray*}
\end{footnotesize}
Therefore, 
\begin{footnotesize}	
\begin{eqnarray*}
\lefteqn{|S|\cdot WCC(\{S_1,S_2\}) =} \\
        & = & |S_1|\cdot WCC(S_1)+|S_2|\cdot WCC(S_2)\\
	& = & \sum_{x\in S_1}WCC(x,S_1)+\sum_{x\in S_2}WCC(x,S_2)\\
	& > & \sum_{x\in S}WCC(x,S).
\end{eqnarray*}
\end{footnotesize}
implies 
\begin{footnotesize}
\begin{eqnarray*}
WCC(\{S_1,S_2\}) & > & \frac{1}{|S|}\sum_{x \in S} WCC(x,S) \\
	& = & WCC(S)= WCC(S_1\cup S_2). \quad \square
	\end{eqnarray*}
\end{footnotesize}	

\section{Proof of Theorem 3}\label{appendix:formal_proof_end}

\begin{proof} 
(i) For the $r-1$ vertices $x\in K_r\setminus \{t\}$, we have
$WCC(x,V)=(vt(x,V)/(n-1)=(r-1)/(n-1)$.  For the vertex $t$, we
have $WCC(v,V)=1$. Finally, for the $s-1$ vertices $x\in K_s\setminus
\{t\}$, we have $WCC(x,V)=(s-1)/(n-1)$. As $n-1=r+s-2$, we obtain the
formula~(\ref{f1}).

(ii) For the $r-1$ vertices $x\in K_r$ we have 
$WCC(x,K_r)=1$. For the vertex $t$, we have
\begin{footnotesize}
\begin{align*}
WCC(x,K_r)
=&\frac{{r-1 \choose 2}}{{r-1\choose 2}
 +{s-1\choose 2}}\cdot\frac{n-1}{r-1+s-1}\\
=&\frac{(r-1)(r-2)}{(r-1)(r-2)+(s-1)(s-2)}.
\end{align*}
\end{footnotesize}
For the $s-1$ vertices $x\in K_s\setminus \{t\}$, we have
\begin{footnotesize}
\begin{align*}
WCC(x,K_s\setminus \{t\})
=\frac{{s-2\choose 2}}{{s-1\choose 2}}\cdot\frac{s-1}{s-1}
=\frac{(s-2)(s-3)}{(s-1)(s-2)}.
\end{align*}   
\end{footnotesize}
This gives the formula~(\ref{f2}).

(iii) For $x\in K_r\setminus \{t\}$, 
\begin{footnotesize}	
\begin{align*}
WCC(x,K_r\setminus \{t\})
=\frac{{r-2\choose 2}}{{r-1\choose 2}}
  \cdot\frac{r-1}{r-1}
=\frac{(r-2)(r-3)}{(r-1)(r-2)};
\end{align*} 
\end{footnotesize}
for vertex $t$, 
\begin{footnotesize}
\begin{align*}
WCC(x,\{t\})
=&\frac{{0\choose 2}}{{n-1\choose 2}}
  \cdot\frac{n-1}{0+r-1+s-1}\\
=&\frac{(1-1)(1-2)}{(r+s-2)(r+s-3)} = 0;
\end{align*}
\end{footnotesize}
for $x\in K_s\setminus \{t\}$,
\begin{footnotesize}    
$$
WCC(x,K_s\setminus \{t\})=\frac{(s-2)(s-3)}{(s-1)(s-2)};
$$
\end{footnotesize}
This implies~(\ref{f3}).

(iv) Define $f_1(r,s)=n\cdot WCC(\mathcal{P}_1)$, $f_2(r,s)=n\cdot
WCC(\mathcal{P}_2)$, and $f_3(r,s)=n\cdot WCC(\mathcal{P}_3)$. The
expression of these functions are those in~(\ref{f1}), (\ref{f2}) and
(\ref{f3}), respectively.  The goal is to show that for all integers
values $r$, $s$ with $r\ge s\ge 4$ the inequality
$f_3(r,s)\le f_2(r,s)$
holds. Clearly, the first summand of $f_3(r,s)$ is smaller
than the first summand of $f_2(r,s)$, and the third summands are
equal. Then, to prove  $f_3(r,s)\le f_2(r,s)$ it is sufficient 
to compare the second summands. As 
the second sumand of $f_3(r,s)=0$, then $f_3(r,s)\le f_2(r,s)$.

(v) We shall prove
$f_2(r,s)-f_1(r,s)\ge 0$ for $n\ge 7$ and $4\le r\le n-3$.
We have $s=n-r+1$ and 
\begin{align*}
f_2(r,s)-f_1(r,s)> & n-4-\frac{(r-1)^2+(n-r)^2}{n-1}\\
 =&\frac{-2r^2+(2+2n)r-5n+3}{n-1}.
\end{align*}
The sign of $f_2(r,s)-f_1(r,s)$ is the sign of the polinomical function
$-2r^2+2(n+1)r-5n+3$, which is a convex function on $r$ with roots:
\newline
\newline
\begin{footnotesize}
$r_1=\frac{1}{2}(n+1-\sqrt{n^2-8n+7})$; 
$r_2=\frac{1}{2}(n+1+\sqrt{n^2-8n+7})$.
\end{footnotesize}
\newline
\newline
Now, for $n\ge 7$, we have $r_1\le 4$ and $r_2\ge n-3$. Therefore, for each
$r\in\{4,\ldots, n-3\}$ we have $f_2(r,s)-f_1(r,s)\ge 0$.\end{proof}

\end{appendix}


\end{document}